\documentclass[]{article}
\pdfoutput=1
\usepackage{amsmath}
\usepackage{cuted}
\usepackage{natbib}  
\usepackage[pdftex]{graphics}
\usepackage{caption}
\usepackage{geometry}
\usepackage{mwe}
\usepackage{float}
\usepackage{mathtools}
\usepackage{breqn}
\usepackage{amsmath}
\usepackage{amssymb}
\usepackage{multirow}
\usepackage{hyperref}
\graphicspath{./images/}
\usepackage{natbib,twoopt}

\DeclarePairedDelimiter{\abs}{\lvert}{\rvert}
\usepackage{lineno}
\DeclareGraphicsExtensions{.eps, .png}
\graphicspath{ {./images/} }
\newcount\colveccount
\newcommand*\colvec[1]{
        \global\colveccount#1
        \begin{pmatrix}
                \colvecnext
        }
        \def\colvecnext#1{
                #1
                \global\advance\colveccount-1
                \ifnum\colveccount>0
                \\
                \expandafter\colvecnext
                \else
        \end{pmatrix}
        \fi
}
\title{Two-body model for the spatial distribution of dust ejected from an atmosphereless body}
\author{Anastasiia Ershova, J\"urgen Schmidt\\
	University of Oulu, Finland}
\begin{document}
\maketitle
\begin{abstract}
	We present a model for the configuration of noninteracting material that is ejected in a continuous manner from an atmosphereless gravitating body for a given distribution of sources. The model is applicable to material on bound or unbound trajectories and to steady and nonsteady modes of ejection. For a jet that is inclined to the surface normal, we related the distributions of ejection direction, velocity, and size to the phase-space number density at the distance from the source body. Integrating over velocity space, we obtained an expression from which we inferred the density, flux, or optical depth of the ejected material. As examples for the application of the code, we calculate profiles of the dust density in the Enceladus plume, the pattern of mass deposition rates around a plume on Europa, and images of optical depth following the nonstationary emission of material from a volcano on Io. We make the source code of a Fortran-95 implementation of the model freely available.
\end{abstract}

\section{Introduction}
The ejection of material from the surfaces of atmosphereless bodies is a ubiquitous phenomenon in the Solar System. Prominent examples are comets, active asteroids, ejecta clouds from hypervelocity impacts, or plumes from active satellites. For the dynamics of the ejected material, it is in many cases possible to neglect any other forces than the mass point gravity of the source body  to a good degree of approximation. This is for instance the case for impact-generated dust clouds around planetary satellites as were detected around the Galilean moons \citep{1999Natur.399..558K, 2003Icar..164..170K} or the Moon \citep{2015Natur.522..324H}, or dust plumes ejected from cryovolcanically active satellites \citep{2006Sci...311.1416S, 2006Sci...311.1393P, 2015GeoRL..4210541S}. We note that for higher-order gravity terms to be negligible, the source body does not necessarily need to be spherical. For instance, mass-point gravity can be a good approximation to describe dust ejection from a satellite with surface topography.

In this paper we derive a semianalytical model to assess the spatial configuration of the emitted dust.  The model relates the distribution of dust sources on the surface of the atmosphereless body and the parameters of ejection (e.g., source strength or directional and velocity distribution) to observable quantities such as number density, fluxes, or optical depth. The mathematical foundations are described in Sect. \ref{sec:mathematics}. Expanding on work in the literature \citep{2003P&SS...51..251K, Sremcevic:2003gf}, our model can handle emission through inclined jets, and we develop a method for carrying out two of three integrations over the velocity distribution analytically. The code that implements the new model, carrying out the one remaining integration numerically, is called DUDI (for \textquotedblleft dust distribution\textquotedblright) and is freely available under the GNU General Public License on https://github.com/Veyza/dudi. Aspects of the numerical algorithm for the integration are outlined in Sect. \ref{sec:numerics}. Examples for an application of the model to current problems in planetary science are given in Sect. \ref{sec:applications}, including cases of steady and nonsteady dust emission.

\section{Mathematical formulation}
\label{sec:mathematics}
\subsection{Phase-space density}
We followed the derivations by \citet{2003P&SS...51..251K}, \citet{Sremcevic:2003gf}, and \citet{2011Natur.474..620P} to relate the phase-space density of dust in a certain point of interest in space to the distributions that describe the ejection of the dust from a source on the moon surface. We generalized the existing model to allow emission from a point source in a direction that is not normal to the surface, with an axisymmetric distribution of ejection angles around this direction. Moreover, we allowed for a general coupling of the distribution of ejection velocities and grain size.

Our model was developed initially to fit in situ measurements by the Cassini Cosmic Dust Analyzer at the Saturn satellite Enceladus. For convenience, we use the words \textquotedblleft spacecraft position\textquotedblright \ or \textquotedblleft spacecraft coordinates\textquotedblright\   from now on to denote the point in space at which the dust density is calculated. We also use the term \textquotedblleft density\textquotedblright,  which at any point can be understood as the number density, mass density, the average radius of dust particles, or the cross section that is covered by the dust at the spacecraft position. The model allows us to obtain any of these quantities with a change of only one parameter.

\begin{figure}
        \centering
        \includegraphics[width=0.5\textwidth]{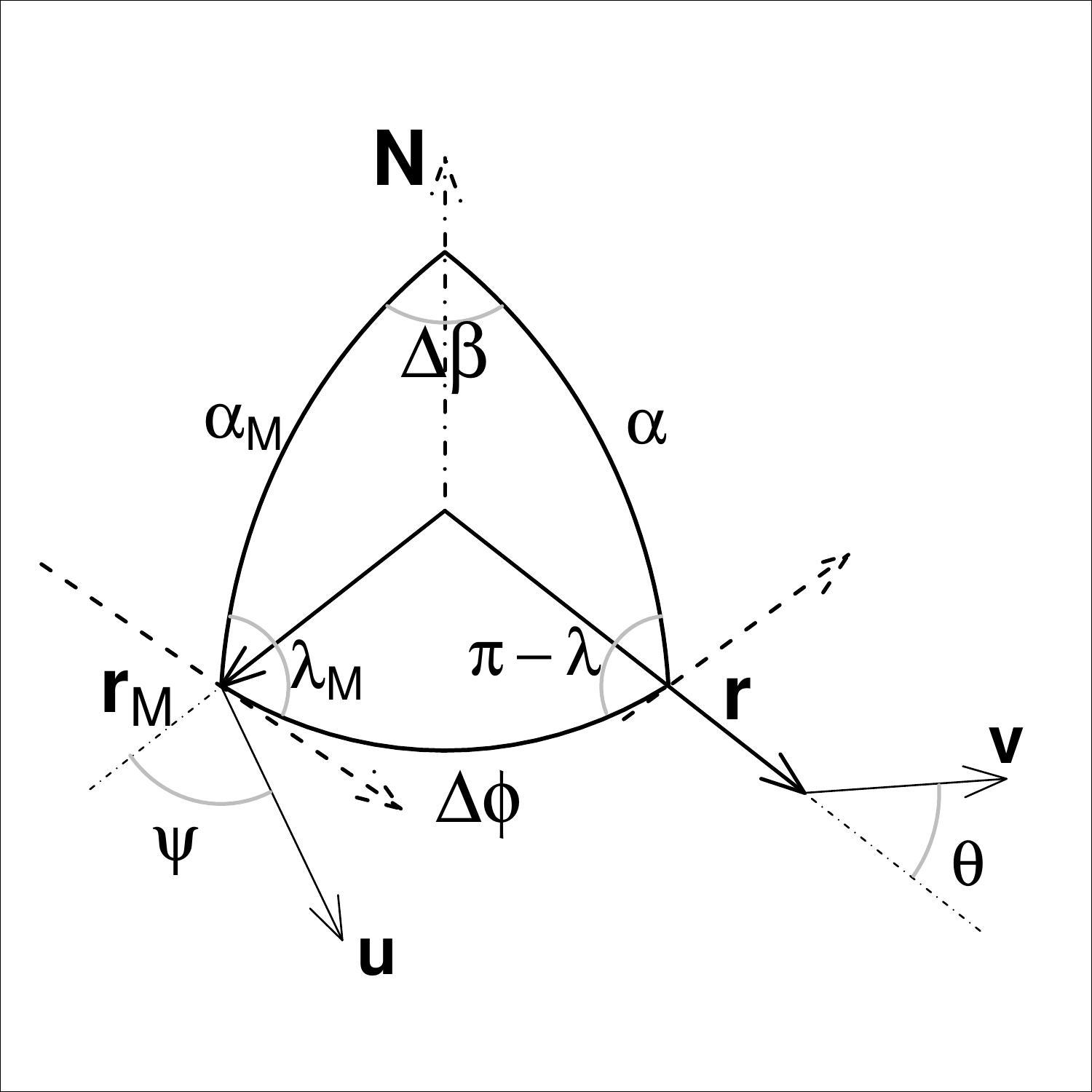}
        \caption{Directions to the north and to the positions of the source and the spacecraft form a spherical triangle. This establishes the relations between the angles of the problem. Here $\Delta\beta$ is the angle between the projections of vectors $\mathbf{r}$ and $\mathbf{r}_M$ on the equatorial plane.}
        \label{sptr}
\end{figure}

We first consider a stationary process. We can equate the differential number of dust particles in a certain point of phase space 
        \begin{equation}
                \label{dninspace}
                \mathrm{d}n = n(r, \alpha, \beta, v, \theta, \lambda,R)r^2\sin\alpha \mathrm{d}r\mathrm{d}\alpha \mathrm{d}\beta\  v^2\sin\theta \mathrm{d}v\mathrm{d}\theta \mathrm{d}\lambda \mathrm{d}R
        \end{equation}
to the number of particles ejected from the satellite surface
        \begin{equation}
                \label{dnonsurface}
                \mathrm{d}n = \gamma \mathrm{d}t\ f(\alpha_M, \beta_M, u, \psi,\lambda_M, R)\sin\alpha_M \mathrm{d}\alpha_M \mathrm{d}\beta_M \mathrm{d}u \sin\psi \mathrm{d}\psi \mathrm{d}\lambda_M \mathrm{d}R
        .\end{equation}
The variables used here are defined in Table \ref{vars}, and Fig. \ref{sptr} illustrates the geometry of the problem.
For the phase-space  density at the spacecraft, we obtain
\begin{multline}
\label{ndens0}
n(r, \alpha, \beta, v, \theta, \lambda, R) v^2 \sin \theta = \frac{\gamma}{|\mathrm{d}r/\mathrm{d}t|r^2}\frac{\sin\alpha_M\sin\psi}{\sin\alpha}\times\\\times f(\alpha_M, \beta_M, u, \psi, \lambda_M, R) \abs*{\frac{\partial (\alpha_M, \beta_M, u, \psi, \lambda_M)}{\partial(\alpha, \beta, v, \theta, \lambda)}}.
\end{multline}
For the two-body problem, the Jacobian can be obtained analytically (see \citet{Sremcevic:2003gf}),
\begin{equation}
\label{Jalpha}
\abs*{\frac{\partial (\alpha_M, \beta_M, u, \psi, \lambda_M)}{\partial(\alpha, \beta, v, \theta, \lambda)}} = \frac{r}{r_M}\frac{v^2}{u^2}\frac{|\cos\theta|}{\cos\psi}\frac{\sin\alpha}{\sin\alpha_M}
.\end{equation}

\begin{table*}[t]
        \caption{Definition of variables}
        \vskip0.3cm
        \begin{tabular}{|c|p{6cm}|c|p{6cm}|}
                \hline
                Variable & Definition & Variable & Definition \\
                \hline
                $r$ & Radial distance from the moon center to a point in space where the density is to be calculated (spacecraft position) & $r_M$ & Radial distance of the dust source from the center of the moon (source position)\\
                $\alpha$ & Colatitude of the spacecraft (measured from the moon north pole) & $\alpha_M$ & Colatitude of the source  (measured from the moon north pole) \\
                $\beta$ & Eastern longitude of the spacecraft & $\beta_M$ & Eastern longitude of the source \\
                $v$ & Particle speed at the spacecraft position & $u$ & Particle speed at the moment of ejection \\
                $\theta$  & Angle
                between the particle velocity and position vectors & $\psi$  & Initial angle
                between the particle velocity and position vectors\\
                $\lambda$ & Azimuth angle of the particle velocity, measured clockwise from local north. & $\lambda_M$ & Azimuth angle of the particle initial velocity, measured clockwise from local north.\\
                $R$ & Radius of the particle & $\gamma$ & Rate of dust particle production \\
                $n$ & Phase-space density of particles with fixed radius & $f$ & Distribution describing the dust ejection process\\
                $\zeta$ & Zenith angle of the source symmetry axis & $\eta$ & Azimuth of the source symmetry axis measured clockwise from local north \\
                 & & $\eta^*$ & Auxiliary angle used in the derivation of the expression for the ejection direction distribution in case of a tilted symmetry axis\\
                \hline
        \end{tabular}
        \label{vars}
\end{table*}
We assume that the distribution function $f$ factorizes, and that the distributions of the source position ($f_{\alpha_M,\beta_M}$), ejection speed ($f_u$), ejection direction ($f_{\psi,\lambda_M}$), and size of the ejected dust particles ($f_R$) can be defined separately,
\begin{equation}
\label{fff}
f(\alpha_M, \beta_M, u, \psi, \lambda_M, R) = f_{\alpha_M,\beta_M} (\alpha_M,\beta_M) f_{\psi, \lambda_M}(\psi, \lambda_M) f_u(u, R)  f_R(R).
\end{equation}

It is physically plausible that the distribution of the ejection speed of the dust particles depends on the grain size \citep{2008Natur.451..685S, 2011Natur.474..620P}, which we emphasize with the notation $f_u(u,R)$.  

We describe the position of the point source located at the coordinates $(\alpha_M^0,\beta_M^0)$ on the surface of the spherical moon as the product of two Dirac $\delta$-functions,
\begin{equation}
\label{fpos}
f_{\alpha_M,\beta_M} (\alpha_M,\beta_M) = \frac{\delta(\alpha_M - \alpha_M^0)\,\delta(\beta_M - \beta_M^0)}{\sin\alpha_M} \equiv\frac{\delta\left((\alpha_M,\beta_M)-(\alpha_M^0,\beta_M^0)\right)}{\sin\alpha_M}
.\end{equation}
The term $\sin\alpha_M$ in the denominator comes from the normalization.

\begin{figure}
        \centering
        \includegraphics[width=0.5\textwidth]{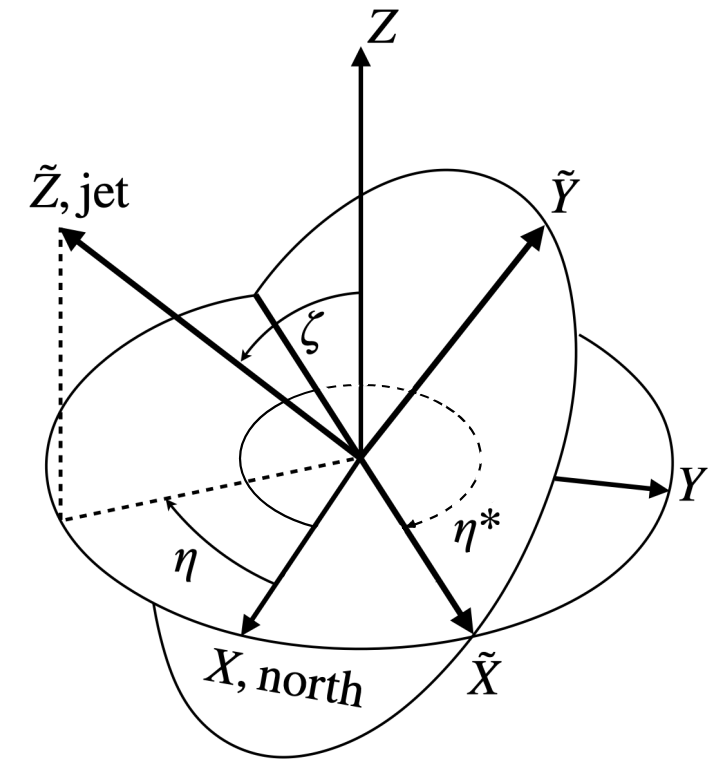}
        \caption{Two coordinate systems centered at the location of the dust source on the surface of a spherical body. The  $Z$-axis of system ($X,Y,Z$) is normal to the surface, and $X$ points to the local north. The $\tilde{Z}$-axis of the coordinate system ($\tilde{X},\tilde{Y}, \tilde{Z}$) is aligned with a jet that is tilted from the surface normal by an angle $\zeta$.}
        \label{2sys}
\end{figure}

To formulate the directional distribution of $\psi$ and $\lambda_M$ so that it describes the ejection of dust that is axisymmetric around the axis of an inclined jet, we consider two coordinate systems centered at the location of the point source. The \textit{Z}-axis of system $(X,Y,Z)$ points along the local normal to the surface, and the $X$-axis points toward the local north. The  $\tilde{Z}$-axis
of system $(\tilde{X}, \tilde{Y}, \tilde{Z})$ is aligned with the axis of the jet. 
 The axis $\tilde{X}$ lies on the line of nodes, so that the angle $\eta^*$ measured from $X$ to $\tilde{X}$ is related to the jet azimuth $\eta$ as $\eta^* = \eta - \pi / 2$. We define azimuth angles always \emph{\textup{clockwise}} from the local north, allowing a direct comparison to the derivations in \citet{2003P&SS...51..251K} and \citet{Sremcevic:2003gf}. Then, the transformation of the $(X,Y,Z)$ coordinate system to the $(\tilde{X}, \tilde{Y}, \tilde{Z})$ coordinate system may be performed as two subsequent rotations, as shown in Fig. \ref{2sys}. The first rotation is clockwise around the $Z$-axis with angle $\eta^*$. The second rotation is counterclockwise around the $\tilde{X}$-axis with angle $\zeta$.

Let $(\psi,\lambda_M)$ and $(\tilde{\psi}, \tilde{\lambda}_M)$ be the polar angle and azimuth in the two systems $(X,Y,Z)$ and $(\tilde{X}, \tilde{Y}, \tilde{Z})$, respectively. The distribution of the ejection direction we wish to use can be formulated in a simple manner in terms of the variables $(\tilde{\psi}, \tilde{\lambda}_M)$ because the distribution is symmetrical with respect to the axis $\tilde{Z}$. 
However, in Eq. (\ref{ndens0}) and in the formulae that are to be derived later in the course of solving the two-body problem, we have to deal with the angles $(\psi,\lambda_M)$ that are defined in the local horizontal coordinate system. Therefore a replacement of the coordinates must be performed in the expression to obtain the distribution of $\psi$ and $\lambda_M$ , which can be used in further calculations. The desired function $f_{\psi, \lambda_M}(\psi, \lambda_M)$ can be obtained by multiplication with the Jacobian,
\begin{equation}
\label{transpsi}
f_{\psi, \lambda_M}(\psi, \lambda_M)\sin\psi = f_{\tilde{\psi}, \tilde{\lambda}_M}(\tilde{\psi}, \tilde{\lambda}_M) \sin\tilde{\psi} \abs*{\frac{\partial (\tilde{\psi}, \tilde{\lambda}_M)}{\partial (\psi, \lambda_M)}}
.\end{equation}
To express $\tilde{\psi}$ and $\tilde{\lambda}_M$ through $\psi$ and $\lambda_M$ , we consider a unit vector $\mathbf{k}$ pointing in an arbitrary direction (Fig. \ref{2sys}). In both coordinate systems the vector can be defined by Cartesian coordinates related to the corresponding polar coordinates as (we recall that $\lambda_M$ and $\tilde{\lambda}_M$ are azimuthal angles counted \emph{\textup{clockwise}} from the $X$ and $\tilde X$ axes, respectively)

\begin{equation}
\label{kvec}
\mathbf{k} = \colvec{3}{k_1}{k_2}{k_3} = \colvec{3}{\sin\psi\cos\lambda_M}{-\sin\psi\sin\lambda_M}{\cos\psi},\ 
\mathbf{k} = \colvec{3}{\tilde{k}_1}{\tilde{k}_2}{\tilde{k}_3} = \colvec{3}{\sin\tilde{\psi}\cos\tilde{\lambda}_M}{-\sin\tilde{\psi}\sin\tilde{\lambda}_M}{\cos\tilde{\psi}}
.\end{equation}
The Cartesian coordinates of $\mathbf{k}$ in the two systems are related through the rotation matrix, which can be expressed in terms of  $\eta$,
\begin{equation}
\label{rotmatr}
\colvec{3}{\tilde{k}_1}{\tilde{k}_2}{\tilde{k}_3} = \begin{pmatrix}
\sin\eta & \cos\eta & 0\\
-\cos\eta\cos \zeta &\sin \eta \cos \zeta & \sin \zeta\\
\cos\eta\sin \zeta& -\sin\eta\sin \zeta & \cos \zeta
\end{pmatrix}\colvec{3}{k_1}{k_2}{k_3}
.\end{equation}
Using Eqs. \ref{kvec} and \ref{rotmatr}, we obtain
\begin{equation}
\label{wpsi}
\tilde{\psi} = \arccos (\cos \zeta\cos\psi + \cos(\eta-\lambda_M)\sin \zeta\sin\psi),
\end{equation}
\begin{equation}
\label{wlambdaM}
 \tilde{\lambda}_M =\arctan \left( \frac{\cos\zeta \sin\psi \cos(\eta-\lambda_M) - \sin\zeta \sin\psi)}{\sin\psi \sin(\eta-\lambda_M)} \right)
.\end{equation}
This gives the Jacobian
\begin{multline}
\label{Jpsi}
\abs*{\frac{\partial (\tilde{\psi}, \tilde{\lambda}_M)}{\partial (\psi, \lambda_M)}} = 4\sin\psi\ / [10-2\cos 2\psi-3\cos 2(\psi - \zeta) - 2\cos 2\zeta -3\cos 2(\psi + \zeta) \\- 8\cos 2(\lambda_M-\eta)\sin^2\zeta\sin^2\psi -8\cos(\lambda_M-\eta)\sin 2\zeta \sin 2\psi] ^{1/2}.
\end{multline}

\subsection{Integration}
\label{integration}
To compute the density of dust at the point $(r, \alpha,\beta),$ we must integrate Eq. (\ref{ndens0}) over all possible velocities and over all possible particle sizes,
\begin{multline}
\label{ndens}
n(r,\alpha,\beta,R_{min} < R < R_{max}) = \frac{\gamma}{r r_M} \int_{v_{min}}^{v_{max}}\mathrm{d}v\int_{0}^{\pi}\mathrm{d}\theta\int_{0}^{2\pi}\mathrm{d}\lambda \frac{v}{u^2}G^p_u(R_{min}, R_{max})\times\\\times\frac{f_{\psi, \lambda_M}(\psi, \lambda_M)\sin\psi }{\cos\psi}\,\frac{\delta\left((\alpha_M(\theta,\lambda),\beta_M(\theta,\lambda))-(\alpha_M^0,\beta_M^0)\right)}{\sin\alpha_M}
.\end{multline}
Here,
\begin{equation}
\label{Gu}
G^p_u(R_{min},R_{max}) \equiv \int_{R_{min}}^{R_{max}}\mathrm{d}Rf_R(R)f_u(u,R)R^p
\end{equation}
is defined in a similar way as in \citet{2011Natur.474..620P}. The parameter $p$ defines the moment of the size distribution related to the quantity we are interested in. Using p=0, we obtain the number density of particles in the specified range of sizes. Setting p=1 gives the average radius of the grains per unit volume. For p=2 we obtain the average cross section of the dust particles per volume. This setting is used below to compute the geometrical optical depth of the dust population. Finally, p=3 gives the average volume occupied by dust grains per unit volume. This setting is used to compute the mass density of the dust. For more details of the evaluation of $G^p_u(R_{min},R_{max}),$ see Appendix B.
We replace variables in the argument of the $\delta$ function in equation (\ref{ndens}) as
\begin{equation}
\label{repl}
\int_{0}^{\pi}\mathrm{d}\theta\int_{0}^{2\pi}\mathrm{d}\lambda\, \delta\left((\alpha_M(\theta,\lambda),\beta_M(\theta,\lambda))-(\alpha_M^0,\beta_M^0)\right)F(\theta,\lambda) =\sum_{i} \frac{F(\theta_i^*,\lambda_i^*)}{\abs*{\frac{\partial(\alpha_M,\beta_M)}{\partial(\theta,\lambda)}}}_{\theta_i^*,\lambda_i^*}\,
.\end{equation}
to integrate over $\theta$ and $\lambda$ analytically.  Eq. (\ref{repl}) is derived in greater detail in Appendix A.
Here, $F(\theta,\lambda)$ represents the integrand of equation \ref{ndens}, while $\theta_i$ and $\lambda_i$ are the roots of the equation
\begin{equation}
\label{alphaMbetaM}
\alpha_M(\theta_i^*,\lambda_i^*) = \alpha_M^0,\quad \beta_M(\theta_i^*,\lambda_i^*)=\beta_M^0
.\end{equation}
All the necessary dependencies between the variables in question can be found from spherical trigonometry  \citep{2003P&SS...51..251K, Sremcevic:2003gf}, for instance,
\begin{equation}
\label{alphaM}
\alpha_M = \arccos\left(\cos\alpha\cos\Delta\phi(\theta) - \sin\alpha\sin\Delta\phi(\theta)\cos\lambda\right)
\end{equation}
and
\begin{equation}
\label{betaM}
\beta_M = \beta \pm \arcsin\left(\frac{\sin\Delta\phi(\theta)\sin\lambda}{\sin\alpha_M(\theta,\lambda)}\right)
.\end{equation}

The spherical triangle used to obtain these relations is shown in Fig. \ref{sptr}. The angle $\Delta\phi$ is the angle between the position vectors of the spacecraft and the source location on the moon.
Because $\theta$ enters expressions \ref{alphaM} and \ref{betaM} only through $\Delta\phi$, the partial derivatives of $\alpha_M$ and $\beta_M$ with respect to $\theta$ can be computed as partial derivatives with respect to $\Delta\phi$ multiplied by $\partial\Delta\phi / \partial\theta$. The Jacobian reads
\begin{equation}
\label{JalphaM}
\abs*{\frac{\partial(\alpha_M,\beta_M)}{\partial(\theta,\lambda)}}=\frac{\sin\Delta\phi}{\sin\alpha_M}\abs*{\frac{\partial\Delta\phi}{\partial\theta}},
\end{equation}
and our final formula is
\begin{multline}
\label{workformula}
n(r,\alpha,\beta,R_{min}<R<R_{max}) = \frac{\gamma}{rr_M\sin\Delta\phi} \int_{v_{min}}^{v_{max}}\mathrm{d}v\frac{v}{u^2}G^p_u(R_{min},R_{max})\times\\ \times \sum_{i}\frac{f_{\psi,\lambda_M}(\psi_i,\lambda_{Mi})\sin\psi_i}{\cos\psi_i}\abs*{\frac{\partial\Delta\phi}{\partial\theta}}_{\theta^*_i}^{-1}.
\end{multline}

The integration over $v$ must be carried out numerically. The lower integration limit is restricted by the minimum energy, or minimum semimajor axis, of the orbits that pass through the two points $(r_M,\alpha_M,\beta_M)$ and $(r,\alpha,\beta)$. It is obtained from
\begin{equation}
        \label{vmin}
        v_{min} = \sqrt{GM\left(\frac{2}{r}-\frac{1}{a_{min}}\right)}
,\end{equation}
where
 \begin{equation}
\label{amin}
a_{min} = \frac{r + r_M}{4} + \frac{1}{2} \sqrt{\frac{r^2 + r_M^2}{4} - \frac{r r_M \cos\Delta\phi}{2}}
.\end{equation}
The upper limit is constrained by the maximum ejection speed,
\begin{equation}
        \label{vmax}
        v_{max} = \sqrt{u_{max}^2+ 2GM \left(\frac{1}{r}-\frac{1}{r_M}\right)}
.\end{equation}

The angle $\Delta\phi$  in the expressions above is the angle between the position vector of the dust source, $\mathbf{r}_M$ , and the position vector of the spacecraft, $\mathbf{r}$. For a fixed position of the source and spacecraft, the angle $\Delta\phi$ is also fixed, but it formally depends on $v$ and $\theta$. In the integrand of equation \ref{workformula}, the value $v$ is given and $\Delta\phi(\theta)$ is a function of the variable $\theta$ alone. From the conservation equations of the two-body problem, we obtain 
\begin{equation}
\label{wp}
\tilde{p} = 2\tilde{r}^2\tilde{v}^2\sin^2\theta,
\end{equation}
\begin{equation}
e^2 = 1 + 4 \tilde{r}^2\tilde{v}^2\sin^2\theta\left(\tilde{v}^2-\frac{1}{\tilde{r}}\right),
\end{equation}
\begin{equation}
\label{cosphiM}
\cos\phi_M= \frac{1}{e}\left(\frac{\tilde{p}}{1}-1\right),
\end{equation}
\begin{equation}
\label{cosphi}
\cos\phi= \frac{1}{e}\left(\frac{\tilde{p}}{\tilde{r}}-1\right),
\end{equation}
\begin{equation}
\label{dphi}
\Delta\phi = \phi - \phi_M,
\end{equation}
which give the relation between $\Delta\phi$ and $\theta$. Following the notational convention of \citet{2003P&SS...51..251K}, we use dimensionless variables $\tilde{r} = r/r_M$ and $\tilde{v} = v/ v_{escape}$, where $v_{escape}$ is the escape velocity on the satellite surface. The angles $\phi$ and $\phi_M$ are the true anomalies at $\mathbf{r}$ and $\mathbf{r}_M$, respectively.

Equations (\ref{wp}) -- (\ref{dphi}) can be used to calculate the partial derivative $\partial\Delta\phi / \partial\theta$. However, it is not possible to reverse these expressions to obtain $\theta$ from a given value of $\Delta\phi$ analytically. The desired $\theta^*_i$ are the values of $\theta$ that (for a given $v$) lead to a $\Delta\phi$ satisfying equations \ref{alphaM} and \ref{betaM}. We use a geometrical approach to calculate all possible $\theta_i^*$. We consider the two-body problem, therefore the motion is restricted to a plane. We define a two-dimensional coordinate system in the plane containing the vectors $\mathbf{r}$ and $\mathbf{r_M}$. The origin is located at the moon center and $\mathbf{r}$ points along the $x$-axis. We know the lengths of $\mathbf{r}$ and $\mathbf{r_M}$ as well as the angle $\Delta\phi$ between them. Then the coordinates of the vectors $\mathbf{r}$ and $\mathbf{r_M}$ in the plane are ($r,0$) and ($r_M\cos\Delta\phi,r_M\sin\Delta\phi$). 

We wish to obtain $\theta$ , which is the angle between $\mathbf{r}$ and the grain velocity vector, which is tangential to the particle trajectory at the point $\mathbf{r}$. This angle can be calculated when we know the equation of the trajectory, which is either an ellipse or a hyperbola. At this step (calculating the integrand of equation \ref{workformula}), we know the value of the particle speed and the distance to the moon center, which determines the orbital energy, and thus, the semimajor axis $a$. We also know whether the particle moves along an ellipse (negative orbital engery) or along a hyperbola (positive orbital energy).

\begin{figure}
        \centering
        \begin{minipage}{0.45\textwidth}
                \centering
                \includegraphics[width=0.9\textwidth]{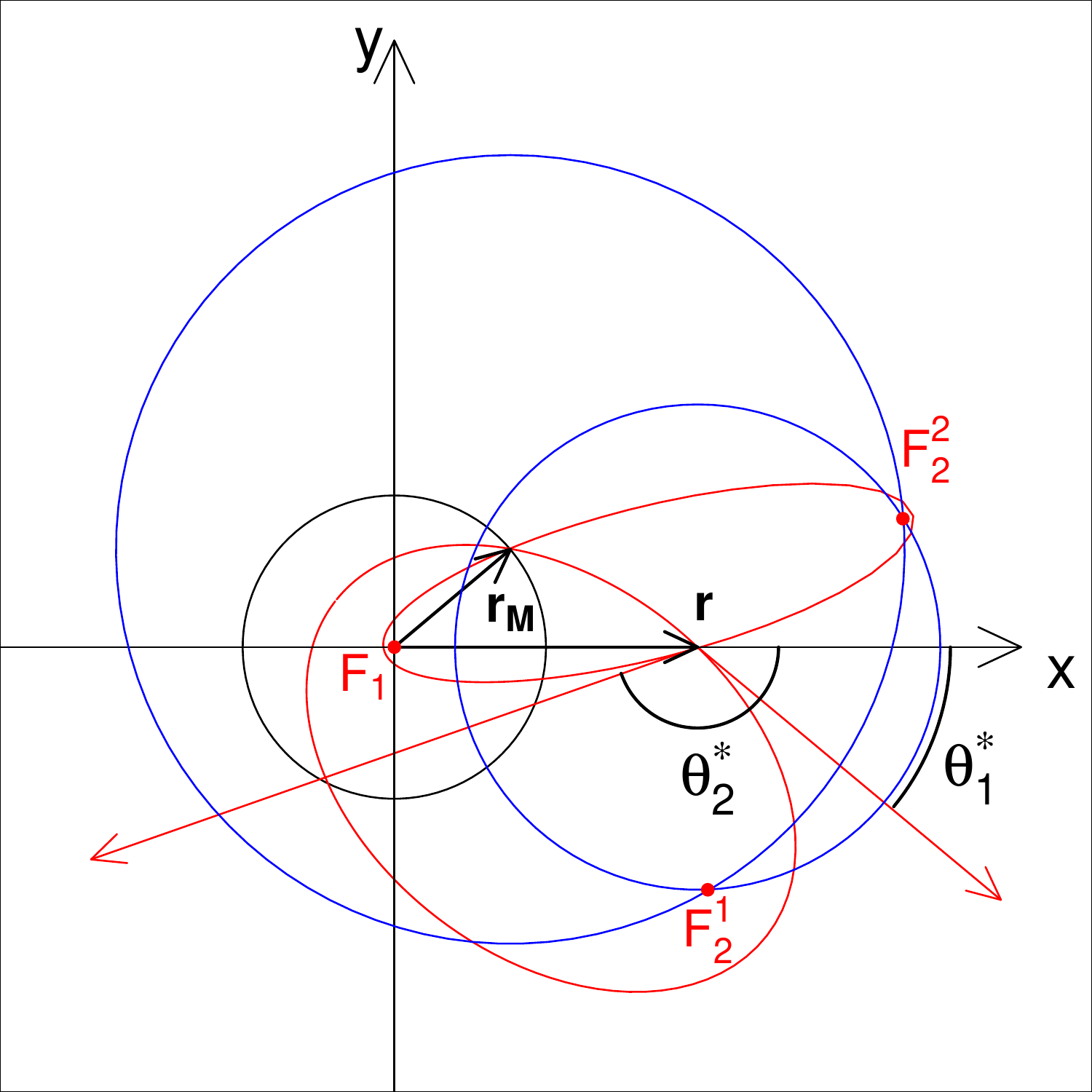} 
                \caption{Finding the second focus for an elliptic trajectory and the solutions for $\theta^*_i$\\
                }
                \label{ellipse}
        \end{minipage}\hfill
        \begin{minipage}{0.45\textwidth}
                \centering
                \includegraphics[width=0.9\textwidth]{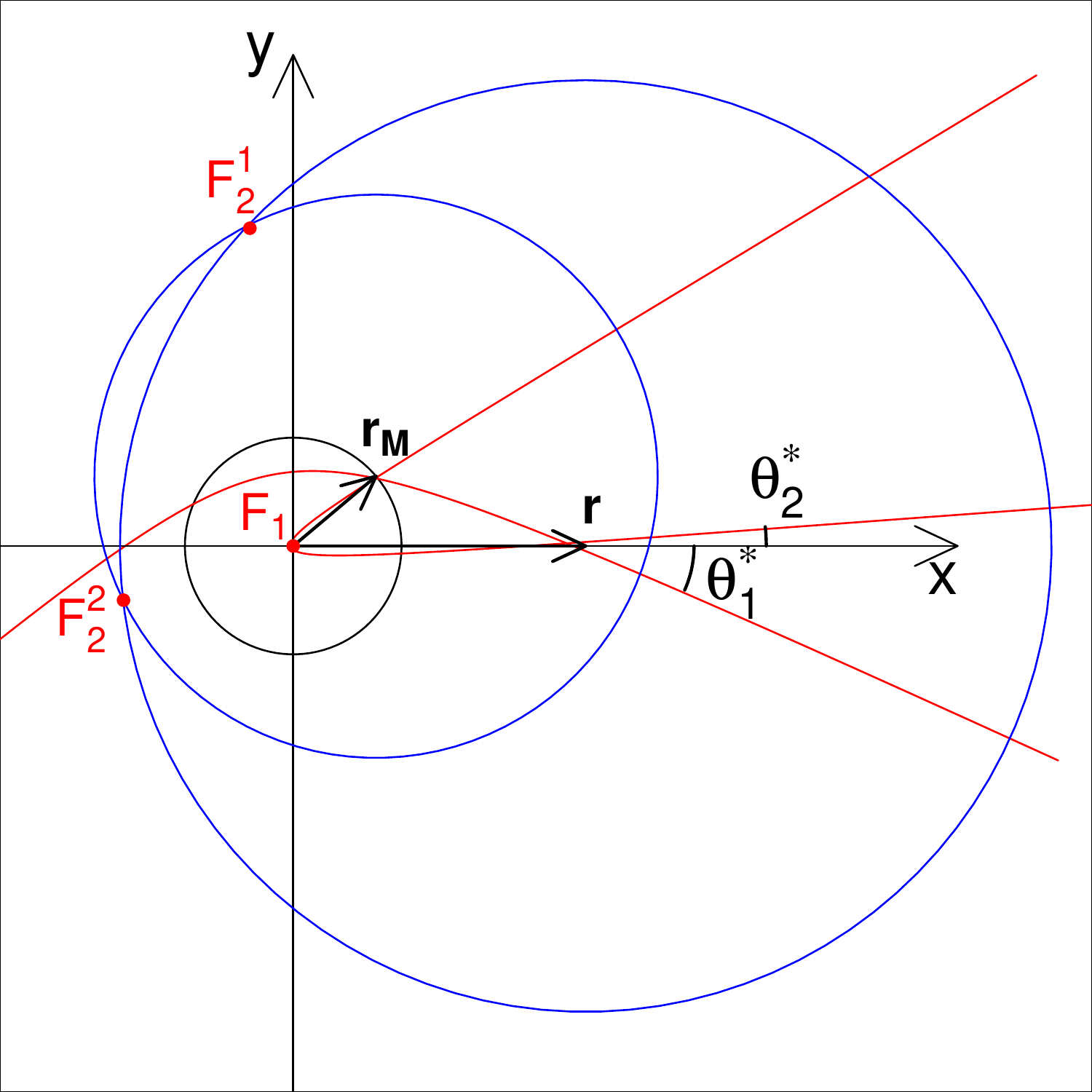} 
                \caption{Finding the second focus for an hyperbolic trajectory and the solutions for $\theta^*_i$. Only $\theta^*_1$ corresponds to a physically possible trajectory.}
                \label{hyperbola}
        \end{minipage}\hfill
\end{figure}

We first consider the case of an ellipse (Fig. \ref{ellipse}). One of the focal points ($F_1$) is located in the origin, which is the center of the moon. To draw the ellipse, we must find the position of the second focus. At each point of the ellipse, the sum of the distances to the focal points is a constant equal to $2a$. Therefore the second focus must be removed from point $\mathbf{r}$ by $2a-r$ and from point $\mathbf{r_M}$ by $2a-r_M$. These conditions are met at the intersection points of two circles centered at $\mathbf{r}$ and $\mathbf{r_M}$ with radii $2a-r$ and $2a-r_M$, respectively.
This leads to the conditions
\begin{equation}
\label{circleel}
\begin{array}{rl}
(x-r_M\cos\Delta\phi)^2 + (y-r_M\sin\Delta\phi)^2 & = (2a-r_M)^2,\\
(x-r)^2 + y^2 &= (2a-r)^2.
\end{array}
\end{equation}
By solving the system of equations \ref{circleel}, we find two possible positions for the second focal point, $F_2^1$ and $F_2^2$. In either case, we can calculate the eccentricity $e$ of the ellipse, the true anomaly at point $\mathbf{r,}$ and, finally, the two solutions for $\theta$ using Eqs. (\ref{eccentricity}) -- (\ref{theta}). Knowing $\cos\phi_M$ is sufficient to obtain $\phi_M$ because we know that the ejection point $\mathbf{r_M}$ cannot have a true anomaly $\phi_M > \pi$. We have
\begin{equation}
\label{eccentricity}
e = \frac{|\overline{F_2F_1}|}{2a},
\end{equation} 
\begin{equation}
\label{cosf1}
        \cos\phi_M = \frac{\overline{F_2F_1}\cdot\mathbf{r}_M}{|\overline{F_2F_1}|r_M},
\end{equation}
\begin{equation}
        \label{difphi}
        \phi = \phi_M + \Delta\phi,
\end{equation}
\begin{equation}
\label{theta}
\theta = \frac{\pi}{2} - \arctan\frac{e\sin \phi}{\sqrt{1+e\cos\phi}}.
\end{equation}

\begin{figure}
        \centering
        \includegraphics[width=0.5\textwidth]{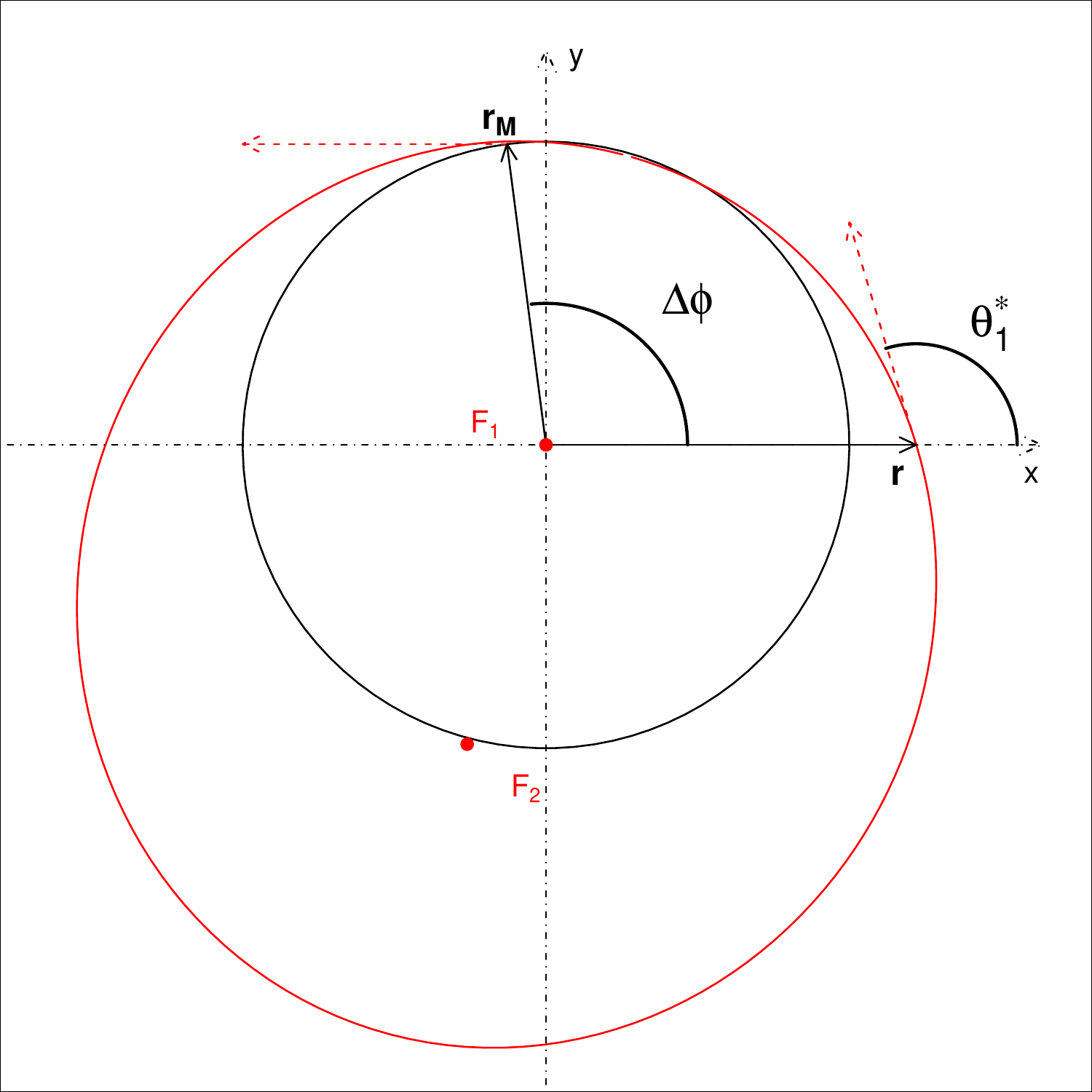}
        \caption{The case when $\Delta\phi$ should be replaced by $2\pi-\Delta\phi.$}
        \label{bigdphi}
\end{figure}

These equations determine the solutions $\theta^*_i$ used in equation \ref{workformula} if the particle travels from $\mathbf{r_M}$ to $\mathbf{r}$ along the shorter arc of the ellipse. However, there are cases when a particle reaches $\mathbf{r}$ over an arc of $2\pi - \Delta\phi$ (Fig. \ref{bigdphi}), leaving $\mathbf{r_M}$ in the opposite direction. To distinguish this case, we must recalculate $r$ from the obtained value for $\phi$ using 
\begin{equation}
\label{rad}
r = \frac{a(1-e^2)}{1+e\cos\phi}
\end{equation}
 and verify that it matches the starting value for $r$ that we used to obtain $\phi$. If this is not the case, then $\Delta\phi$ in Eq. (\ref{difphi}) must be replaced by $2\pi-\Delta\phi$, which corresponds to the motion along the same ellipse, but in the opposite direction. This case is relatively rare, and in the examples we explored was only encountered at large distances from the source.

In case of hyperbolic motion ($a<0$, Fig. \ref{hyperbola}), we follow almost the same line. For every point on a hyperbola, the difference between the distances to the focal points is the same. Because $r_M$ and $r$ must be distances to the nearest focus, the system of equations for the coordinates of the second focal point reads 
\begin{equation}
\label{circlehy}
\begin{array}{rl}
(x-r_M\cos\Delta\phi)^2 + (y-r_M\sin\Delta\phi)^2 & = (r_M+2|a|)^2,\\
(x-r)^2 + y^2 &= (r+2|a|)^2.
\end{array}
\end{equation}
\vskip0.4cm
Equations \ref{eccentricity} -- \ref{theta} remain the same for the hyperbolic case. However, when the coordinates of the second focus are found, we must make sure that the particle does not pass the pericenter on its way from $\mathbf{r_M}$ to $\mathbf{r}$. For this purpose, we verify that the points $\mathbf{r_M}$ and $\mathbf{r}$ lie on the same side of the line $F_1F_2$. If this condition is not satisfied, the solution is rejected. In Fig. \ref{hyperbola} the hyperbola with second focus at the point $F_2^2$  does not meet this condition. Therefore only one hyperbolic trajectory is possible to get from $\mathbf{r_M}$ to $\mathbf{r}$. Furthermore, the motion along the hyperbola is possible only in one direction. The value for $\sin\phi$ is always positive and $\Delta\phi=\phi-\phi_M$.

The values of $\lambda_i$ can be inferred from spherical trigonometry. The two solutions can be either identical or they differ by $180^\circ$ because the motion is restricted to a plane,
\begin{equation}
\label{coslam}
\cos\lambda= \frac{\cos\alpha_M\cos\Delta\phi-\cos\alpha}{\sin\alpha_M\sin\Delta\phi},
\end{equation}
\begin{equation}
\label{sinlam}
\sin\lambda=\pm\frac{\sin\alpha_M\sin(\beta-\beta_M)}{\sin\Delta\phi}
.\end{equation}
The sign in Eq. (\ref{sinlam}) depends on the specific orientation of $\mathbf{r}$ and $\mathbf{r_M}$ relative to the direction of zero-longitude. When the particle travels from $\mathbf{r_M}$ to $\mathbf{r}$ over the angle of $2\pi-\Delta\phi$ , the signs of $\sin\lambda$ and $\cos\lambda$ both change because the sign of the $\sin\Delta\phi$ term in the denominator changes.

As soon as the values of $(v,\theta^*_i,\lambda_i)$ that satisfy the orbital geometry are known, they may be used to calculate the corresponding $(u,\psi_i,\lambda_{Mi})$ from Eqs. (\ref{ufromv}) -- (\ref{coslambdaM}) and the integrand in Eq. (\ref{workformula}) is fully determined,
\begin{equation}
\label{ufromv}
u = \sqrt{v^2_{escape}+2\left(\frac{v^2}{2}-\frac{GM}{r}\right)},
\end{equation}
\begin{equation}
\label{psi}
\sin\psi = \frac{rv}{r_M u}\sin\theta,
\end{equation}
\begin{equation}
\label{sinlambdaM}
\sin\lambda_M=\frac{\sin\alpha\sin\lambda}{\sin\alpha_M},
\end{equation}
\begin{equation}
\label{coslambdaM}
\cos\lambda_M = \frac{\cos\alpha -\cos\alpha_M\cos\Delta\phi}{\sin\alpha_M\sin\Delta\phi}
.\end{equation}

\subsection{Nonstationary case}
Nonstationary dust ejection can be modeled by allowing a time-dependent production rate $\gamma$  in Eq. (\ref{workformula})
that will result in a time-dependent spatial distribution of the dust. When we determine the orbital geometry for a fixed velocity value at the given point in space (Sect. \ref{integration}), the time $\Delta t$ required for traveling from $\mathbf{r_{M}}$ to $\mathbf{r}$ along the Keplerian orbit can be calculated from Kepler's equation. Thus, we know that the properties of the dust configuration at location $\mathbf{r}$ and time $t$ are caused by the production of dust at the source location $\mathbf{r_{M}}$ with the rate $\gamma(t-\Delta t)$.
In this case, the production rate cannot be put outside the integral and Eq. (\ref{workformula}) becomes
\begin{multline}
\label{timedep}
        n(r,\alpha,\beta,R_{min}<R<R_{max},t) = \frac{1}{rr_M\sin\Delta\phi} \int_{v_{min}}^{v_{max}}\mathrm{d}v\frac{v}{u^2}G^p_u(R_{min},R_{max})\times\\\times\sum_{i}\gamma(t-\Delta t_i)\frac{f_{\psi,\lambda_M}(\psi_i,\lambda_{Mi})\sin\psi_i}{\cos\psi_i}\abs*{\frac{\partial\Delta\phi}{\partial\theta}}_{\theta^*_i}^{-1}.
\end{multline}
The two solutions for $\Delta t_i$ follow from Kepler's equation using the two solutions for eccentricity from equation \ref{eccentricity} and computing the eccentric anomaly with the half-angle formula from the true anomaly given by equation \ref{difphi}.

\subsection{Singularities of the coordinates}
\label{spepoints}
There are coordinates for which solutions of equations (\ref{workformula}) or (\ref{timedep}) cannot be obtained (see Table \ref{sppoints}). The problem arises from the use of a spherical coordinate system. In practice, it is possible to avoid the singularities by carefully choosing the pole axis of the coordinate system after the source location and the detector position of interest are known. If coordinates close to the singularities need to be evaluated, then the stability and accuracy of the numerical integration becomes a challenge. However, double-precision calculations allow approaching the singular angles as close as $10^{-4}$ radians. This difference from the values listed in Table \ref{sppoints} can be considered safe, and with a possible moderate loss of accuracy, the model can be applied within an even closer vicinity of the singularities.
\begin{table*}
        \caption{Coordinate singularities of the model}
        \begin{tabular}{|c|c|}
                \hline
                Variable value    &  Physical and geometrical meaning   \\    
                \hline
                $\alpha_M = 0^\circ$ & Point source of dust is located at the north or south pole,  \\
                $\alpha_M = 180^\circ$ & its latitude is not defined, and no spherical triangle from Fig. \ref{sptr} exists\\
                \hline
                $\alpha = 0^\circ$ & Spacecraft is located directly above the north or south pole, \\
                $\alpha = 180^\circ$ & its latitude  is not defined, no spherical triangle from Fig. \ref{sptr} exists\\
                \hline
                $\beta = \beta_M$ & Spacecraft has the same longitude \\
                ($\Delta\beta = 0^\circ$) &as the source. No spherical triangle from Fig. \ref{sptr} exists\\
                \hline
                $\Delta\phi = 0^\circ$& Spacecraft position, source position, and moon center \\
                $\Delta\phi = 180^\circ$ & are on a straight line. No spherical triangle from Fig. \ref{sptr} exists\\
                &  and the orbit geometry (Figs. \ref{ellipse} and \ref{hyperbola}) is undefined\\
                \hline
                $\psi = z$, & A particle is ejected exactly along\\
                $z \neq 0^\circ$ &  the jet axis, the azimuth of ejection is undefined, and the Jacobian\\
                &  in Eq. (\ref{Jpsi}) diverges\\
                \hline
        \end{tabular}
        \label{sppoints}
\end{table*}

\section{Numerical integration}
\label{sec:numerics}
In this section we present and discuss the algorithm for the numerical solution of equations \ref{workformula} and \ref{timedep}. We have implemented this algorithm in a code written in Fortran-95, which we call dust distribution (DUDI). The source code with technical documentation and instructions for usage and compilation is freely available under the GNU General Public License on https://github.com/Veyza/dudi. 
The \textit{makefile} provided for compilation uses the \textit{gfortran} compiler. DUDI can be compiled and run without installing additional libraries. The library of \textit{OpenMP}, which is used to speed the computation up, is included in the compiler. 

DUDI allows us to compute the number density of dust or related quantities at given points in space as a mean radius,  average cross section, or mass density of the dust grains ejected from the surface of a spherical body without an atmosphere. The input data are the spacecraft coordinates, the properties of the source (i.e., the\ location and the distributions of the direction and speed of the ejection) and the three parameters of $G^p_u$. 

The first preliminary step is to calculate $G^p_u$ on a grid of $u$-values. The array of pairs $(u, G^p_u)$ is later used to interpolate $G^p_u(R_{min}, R_{max})$ for the actually required value $u$ under the integrand of equations \ref{workformula} or \ref{timedep}. The second preliminary step is to compute the values of $\Delta\phi$ and $\Delta\beta$ for the given positions of the source and the spacecraft.

Then we proceed directly to the numerical evaluation of the integral over velocity in Eq (\ref{workformula}) in the stationary case, or Eq. (\ref{timedep}) if there is a time-dependence. The ejection speed distribution implies a certain lower limit for the possible ejection speed $u_{min}$. At any point $\mathbf{r,}$ this minimum ejection velocity restricts the corresponding minimum velocity at spacecraft position $v_{min}^0$ , which may be higher than the lower integration limit given by Eqs. (\ref{amin}) -- (\ref{vmin}). The actual numerical integration is performed over the interval where $(v_{min},v_{max})$ and $(v_{min}^0,v_{max})$ overlap. The minimum and maximum ejection speed $u_{min}$ and $u_{max}$ (in m/s) must be explicitly specified as a property of the dust source, along with the expression for the ejection speed distribution.

At each step of the integration, we find for a given $v$, $r$, and $\Delta\phi$ the solutions for the angles $\theta$ and $\lambda$ as described in Sect. \ref{integration}. We control the accuracy of the solution for $\theta$ by recalculating the value of $\Delta\phi$ from Eqs. (\ref{wp})--(\ref{dphi}) and comparing it to the starting value of $\Delta\phi$ as given by the positions of the dust source and the spacecraft. We require the difference between the two values of $\Delta\phi$ to be smaller than $10^{-4}\Delta\phi$. In most cases the accuracy is much better, but it may degrade for certain values of $v$ near the singularities of the coordinates (see Sect. \ref{spepoints}) or when $r \approx r_M$. Even for poor accuracy for $\theta$ , this means that the accuracy decreases only for one or two integration steps. The error in the final result is smaller. We consider the relative accuracy of $10^{-4}$ sufficient for the $\theta$ solutions. If it is worse, a warning is produced by the program.

 We divide the integration domain into two regions as $(v_{min},v_{par})$ and $(v_{par},v_{max})$ to obtain the number density of the particles on elliptic and hyperbolic (escaping) trajectories separately.  Here $v_{par}$ (the subscript stands for "\emph{\textup{parabolic"}}) is the minimum escape velocity at radial distance $r$. 

 When $v_{min} > v_{min}^0$ , the integrand in Eq. (\ref{workformula}) has a pole at $v = v_{min}$. We replace $v_{min}$ from Eq. (\ref{vmin}) with $v_{min} + \Delta$, where $\Delta = 10^{-10}$ has turned out to be a good choice to evaluate the integrand near the pole to reasonable accuracy in a stable manner. To better resolve the pole, we additionally subdivide the elliptic part of the integral into two parts that are treated separately. The first integration subinterval is $(v_{min},v_{1})$, where $v_{1} = v_{min} + 10^{-4}(v_{par} - v_{min})$ is the domain that contains the pole. Here integration is performed using the trapezoidal rule with a large number of supports that become denser toward $v_{min}$ as
 \begin{equation}
 \label{vsteps}
 v_i = v_{min} + \left(\frac{i}{N}\right)^k (v_{1} - v_{min})
 ,\end{equation}
 where $N$ is the number of supports. Then we use the Gauss-Legendre quadrature of a moderate order to compute the integral from $v_{1}$ to $v_{par}$. The integration over the hyperbolic velocities is also performed with a Gauss-Legendre quadrature. The choice of the quadrature order depends on the integration domain and ejection speed distribution. The nodes and weights of the Gauss-Legendre formula are tabulated in our code for the following orders: 5, 10, 20, 30, 40, and 50. 
 
\begin{figure}
        \centering
        \begin{minipage}{0.45\textwidth}
                \centering
                \includegraphics[width=0.9\textwidth]{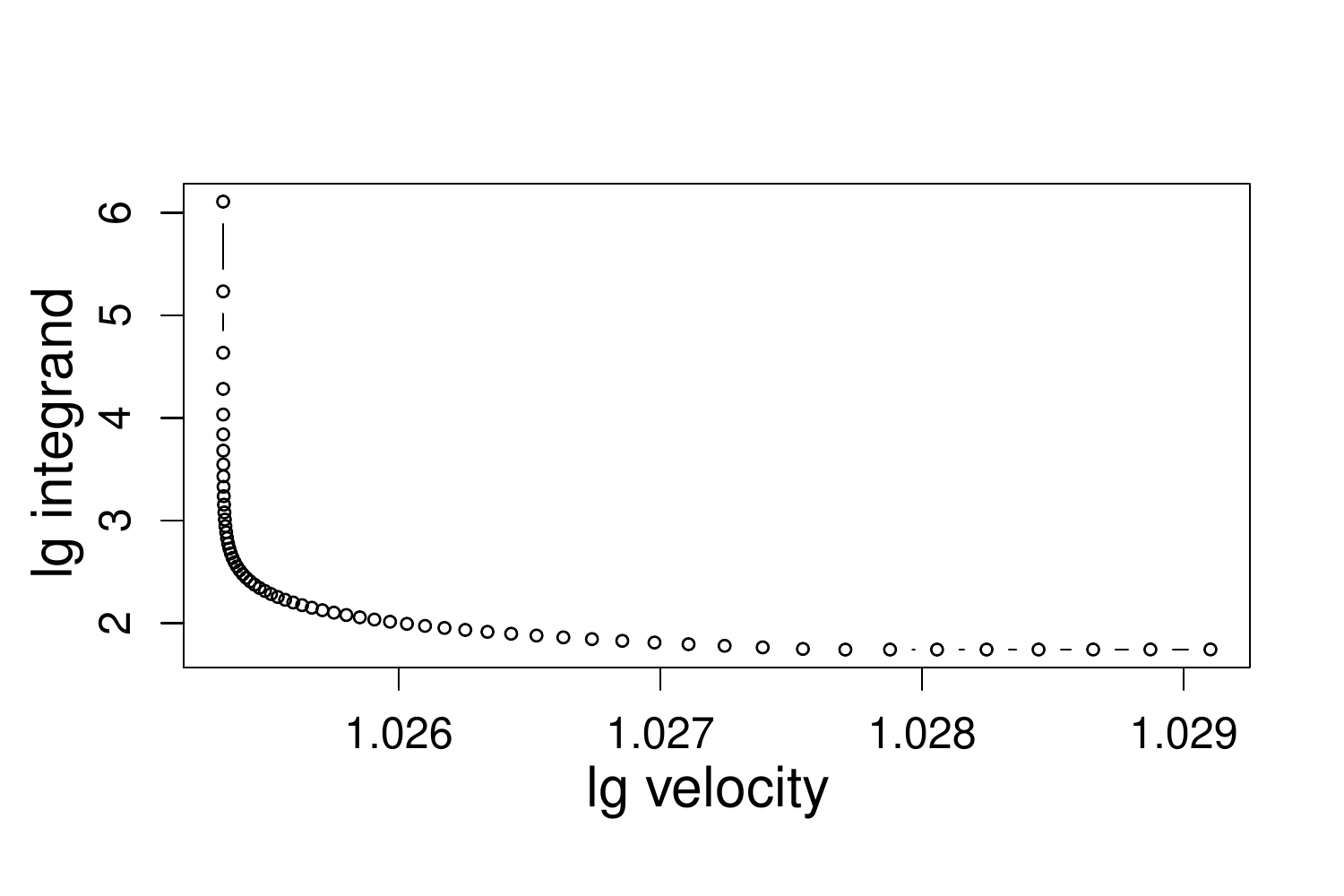} 
                \caption{Pole at $v = v_{min.}$\\
                }
                \label{integrandpole}
        \end{minipage}\hfill
        \begin{minipage}{0.45\textwidth}
                \centering
                \includegraphics[width=0.9\textwidth]{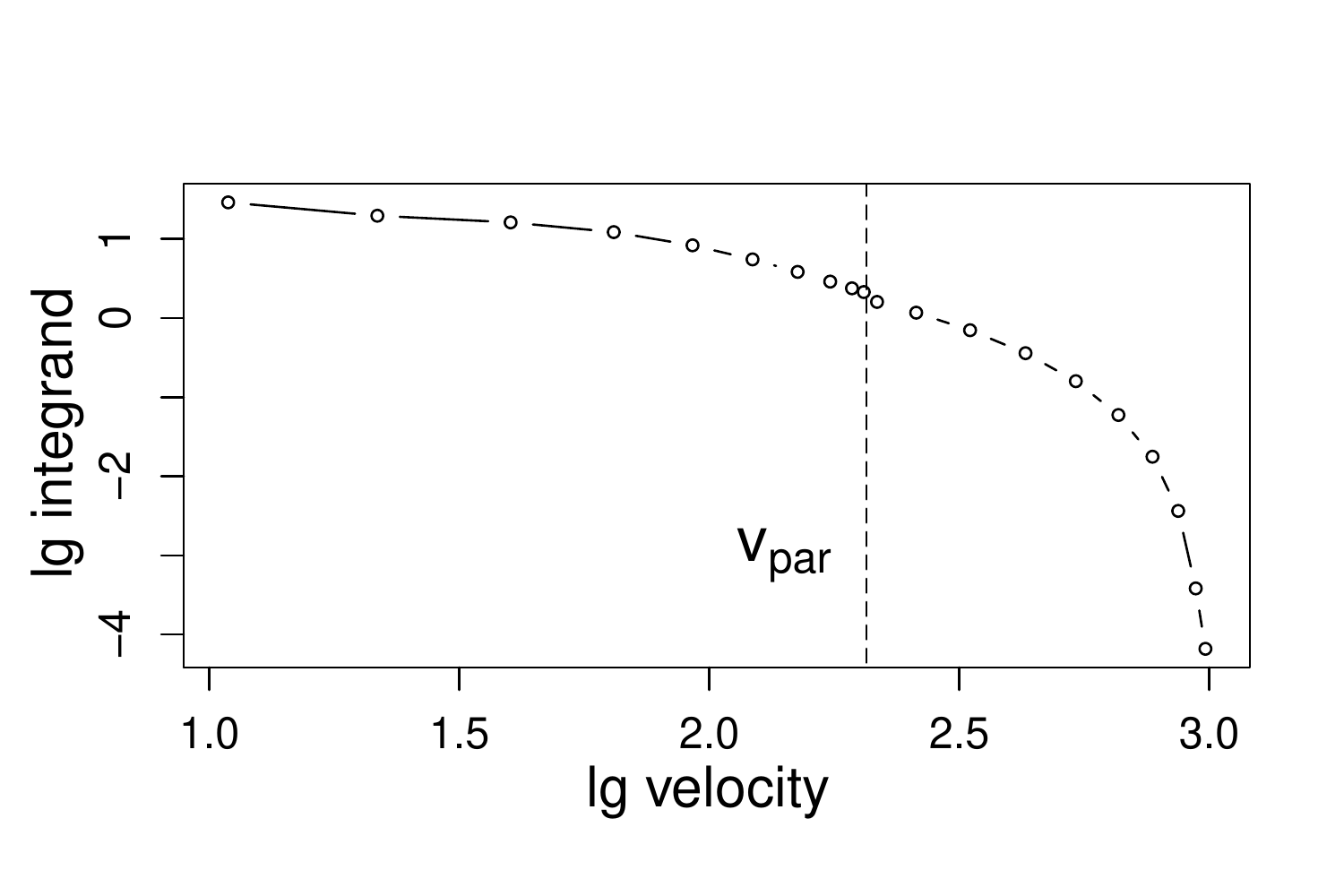}
                \caption{Integrand from Eq. (\ref{workformula}) for an ejection speed distribution that favors low velocities.}
                \label{integrand1}
        \end{minipage}\hfill\\
        \begin{minipage}{0.45\textwidth}
                \centering
                \includegraphics[width=0.9\textwidth]{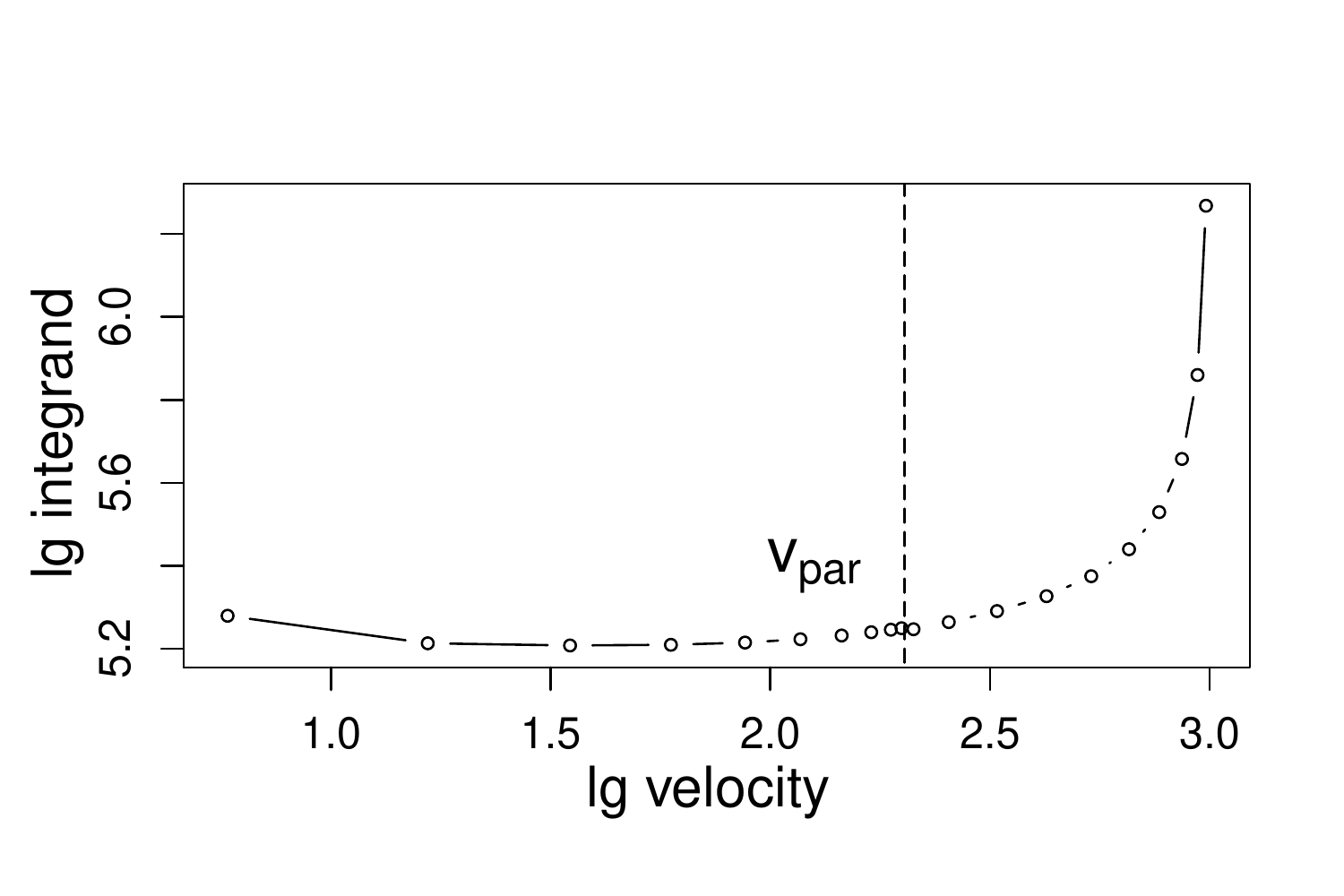} 
                \caption{Integrand from Eq. (\ref{workformula}) for an ejection speed distribution that favors high velocities.\\
                }
                \label{integrand2}
        \end{minipage}\hfill
        \begin{minipage}{0.45\textwidth}
                \centering
                \includegraphics[width=0.9\textwidth]{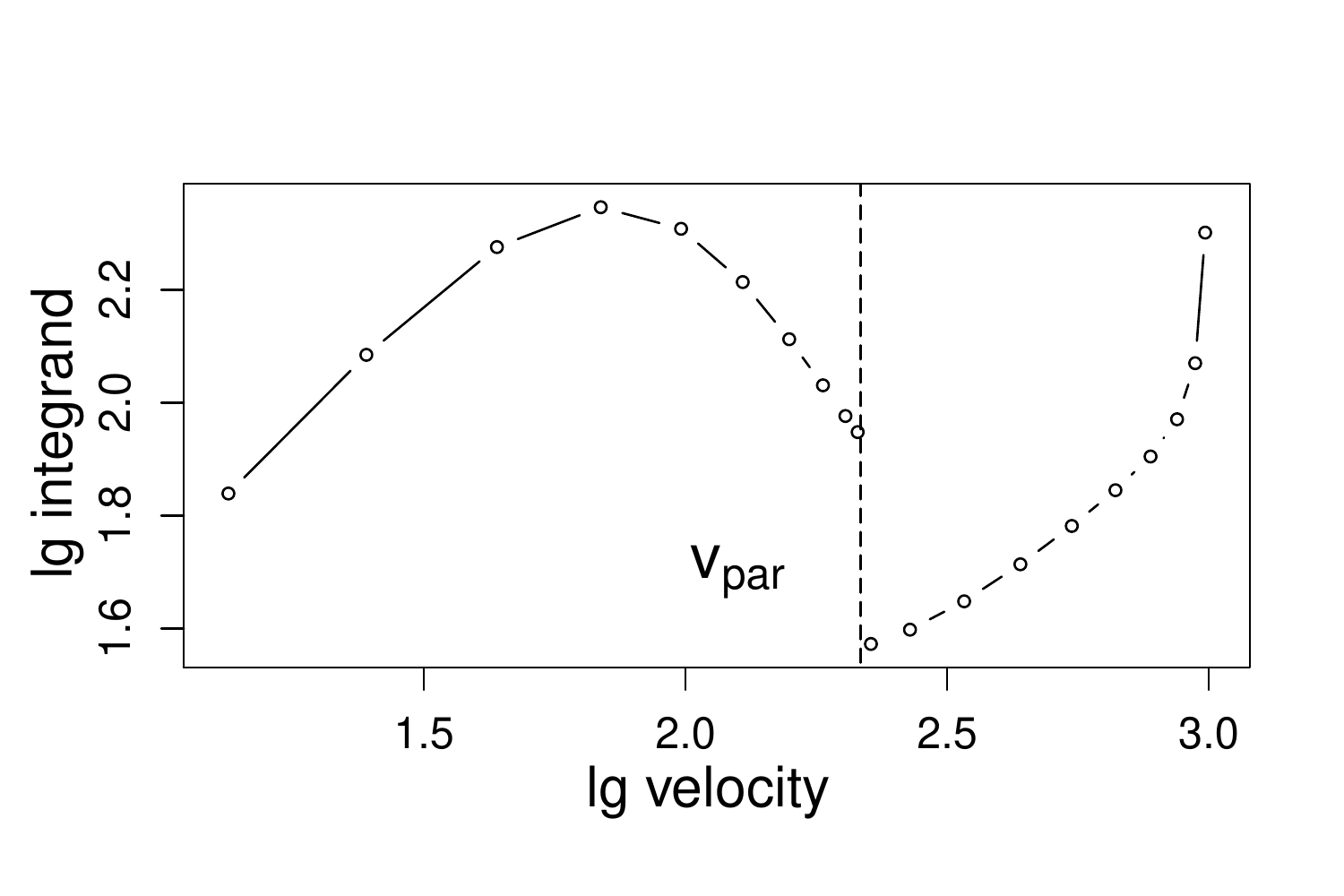} 
                \caption{Integrand from Eq. (\ref{workformula}). A high abundance of dust falling back causes a jump at the transition from the elliptic to the hyperbolic case.}
                \label{integrand3}
        \end{minipage}\hfill
        \captionsetup{labelformat=empty}
\end{figure}

 Figures \ref{integrandpole} -- \ref{integrand3} show examples for the behavior of the integrand in the three domains. Depending on the choice of the ejection speed distribution $f_u(u,R),$ the integrand may decrease (Fig. \ref{integrand1}) or increase (Fig. \ref{integrand2}) toward higher velocities. Remarkably, the integrand can jump at $v = v_{par}$ (Fig. \ref{integrand3}) if a significant part of the dust number density is due to the particles on their way back to the moon after passage of their apocenter,

 The sharpness of the pole varies. At each integration step, the given velocity $v$ determines the values of $\theta$ that enter the integrand of equation \ref{workformula} through the derivative $\partial\Delta\phi / \partial\theta$ and $\psi(\theta)$. The latter is needed to compute the ejection angle distribution $f_{\psi,\lambda_M}(\psi,\lambda_M)$. The factor $\partial\Delta\phi / \partial\theta$ is the reason for the pole. Its value depends on $\theta$ and on the spacecraft position relative to the source. The pole is less strongly peaked if the value of $\psi(\theta)$ corresponds to a very unlikely ejection direction. Thus, the sharpness of the pole depends on the spacecraft position relative to the source position and also on the ejection angle distribution, and so does the number of integration steps required to achieve a given accuracy goal.

The accuracy can be estimated from Eq. (\ref{eps}), where $P$ is the value of the pole integral (between $v_{min}$ and $v_1$) obtained with $N$ steps for the integration with the trapezoidal rule, and $N_{max}$ is the maximum reasonable number of steps. $N_{max}$ is limited by accumulated rounding errors, and we determine its value in test integrations. $I(N_{max})$ is the sum of $P(N_{max})$ and the remaining part of the integral between $v_1$ and $v_{max}$. In this way, we can quantify the discrepancy induced by the pole integration in the final result,

\begin{equation}
        \label{eps}
        \epsilon(N) = |P(N) - P(N_{max})| / I(N_{max})
.\end{equation}

We require $\epsilon \le 10^{-3}$ and perform tests to determine the corresponding number of pole integration steps $N$ necessary to achieve this goal. This number we compute for different spacecraft positions relative to the source and to the axis of ejection symmetry (the polar angle in the coordinate system $\tilde{X} \tilde{Y} \tilde{Z}$ in Fig. \ref{2sys}, in the following denoted by $\xi$). We adopt 
\begin{equation}
\label{psidistr}
f_{\tilde{\psi}, \tilde{\lambda}_M}(\tilde{\psi}, \tilde{\lambda}_M)\sin\tilde{\psi} = e^{-(\tilde{\psi}-\psi_{max})^2/2\omega^2} \frac{\sin\tilde{\psi}}{2\pi}
.\end{equation}
for the ejection direction distribution. Normalization in this expression does not matter for an evaluation of $\epsilon$ from Eq. (\ref{eps}). We vary the parameters $\psi_{max}$ and $\omega$, along with the polar angle of the ejection symmetry axis, to investigate the behavior of the pole for different ejection distributions, of which two main classes can be defined. The \textquotedblleft jets\textquotedblright\ are the sources with a preferred direction of ejection, and the \textquotedblleft diffuse sources\textquotedblright \ have no such direction. 
 We find that for diffuse ejection ($\omega = 45^\circ$ and $\psi_{max} = 45^\circ$), N = 15 is a sufficient number of supports to achieve $\epsilon \le 10^{-3}$ at any spacecraft position. For a vertical jet ($\psi_{max} = 0^\circ$ and $\omega$ in the range of $3^\circ$ and $5^\circ$), the value of $P$ can be neglected for all $\xi > 40^\circ$ and $N = 15$ is sufficient for $\xi < 40^\circ$.
 
 However, for a narrow and inclined jet, we find that there are points where a large number of supports is required to integrate the pole accurately. The narrower and the more inclined the jet, the greater the number of these points and the greater the required $N$. We focus on the worst-case scenarios generally to constrain an optimal number of steps required for the pole integration.

\begin{figure}[H]
        \centering
        \begin{minipage}{0.45\textwidth}
                \centering
                \includegraphics[scale=0.5]{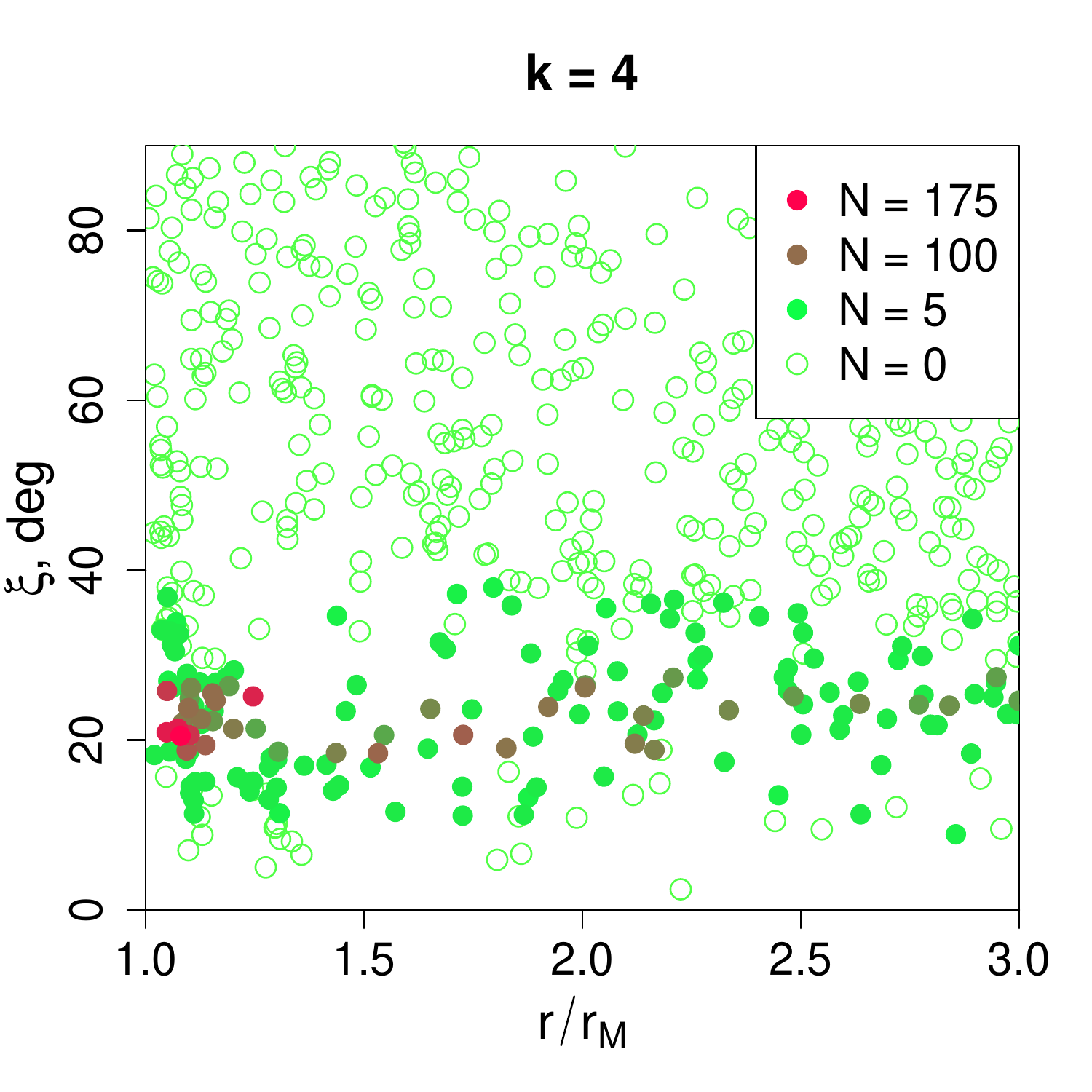}
                \caption{Minimum number of supports necessary to achieve an accuracy of $10^{-3}$ (equation \ref{eps}) in the integration of the pole of the integrand (Fig.\ref{integrandpole})  for the case of a narrow jet ($\omega = 3^\circ,\ \psi_{max} = 0^\circ$), tilted by $z = 20^\circ.$}
                \label{accumap1}
        \end{minipage}\hfill
        \begin{minipage}{0.45\textwidth}
                \centering
                \includegraphics[scale=0.5]{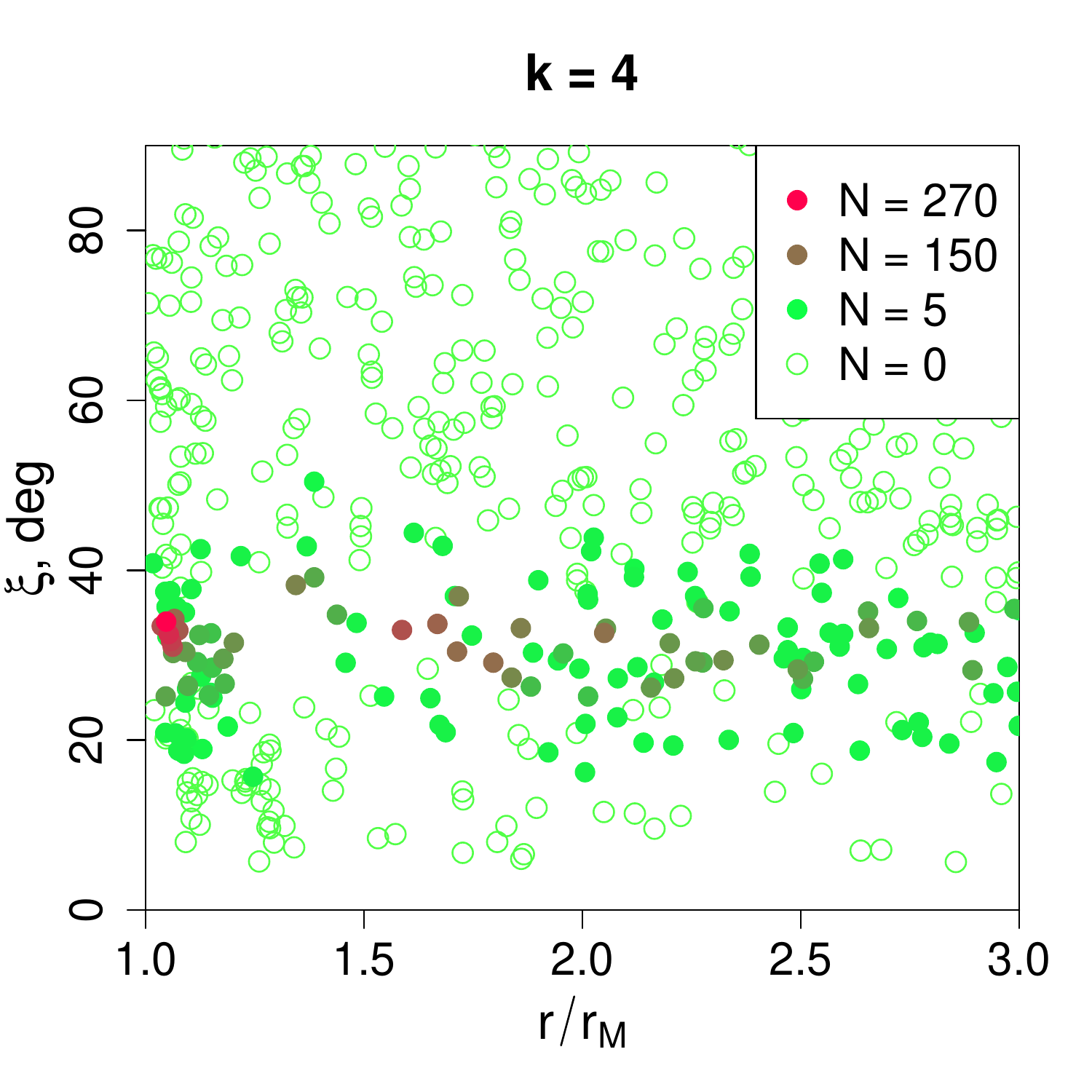}
                \caption{Minimum number of supports necessary to achieve an accuracy of $10^{-3}$ (equation \ref{eps}) in the integration of the pole of the integrand (Fig. \ref{integrandpole})  for the case of a narrow jet ($\omega = 3^\circ,\ \psi_{max} = 0^\circ$), tilted by $z = 30^\circ.$}
                \label{accumap2}
        \end{minipage}\hfill
        
\end{figure}

Empirically, we find that an accuracy of $10^{-3}$ can be achieved with a minimum number of steps when an exponent of $k = 4$ is used in Eq. (\ref{vsteps}). Figs. \ref{accumap1} and \ref{accumap2} show examples of how the minimum number of steps required to integrate the pole with the given accuracy is distributed over $r$ and $\xi$ values. $N=0$ means that the pole does not have to be integrated at all because its value is negligible.

For jets, we select the number of supports $N$ for the pole integration
based on the distributions shown in Figs. \ref{accumap1} and \ref{accumap2} as follows: 
\begin{equation}
\label{Nprestep}
N = \left\{ \begin{array}{cl}
0,& \xi > 45^\circ\ \mathrm{or}\ \xi < 10^\circ,\ r/r_M > 1.05,\\
80,&\xi > 45^\circ\ \mathrm{or}\ \xi < 10^\circ,\ r/r_M < 1.05,\\
15 + 10z[\mathrm{deg}],& 10^\circ < \xi < 45^\circ, \ r/r_M <2,\\
10 + 5z[\mathrm{deg}],& 10^\circ < \xi < 45^\circ, \ r/r_M >2.\\
\end{array}\right.
\end{equation}

\section{Applications}
\label{sec:applications}
In this section we present three applications of the model to phenomena of scientific interest in the Solar System. The purpose of these examples is to demonstrate the wide range of applicability of the model. We leave a rigorous scientific analysis of these problems with a comprehensive comparison to data for future work.

\subsection{Density profile of the Enceladus dust plume}
\label{enceladusex}
\begin{figure}[H]
        \centering
        \includegraphics[width=0.5\textwidth]{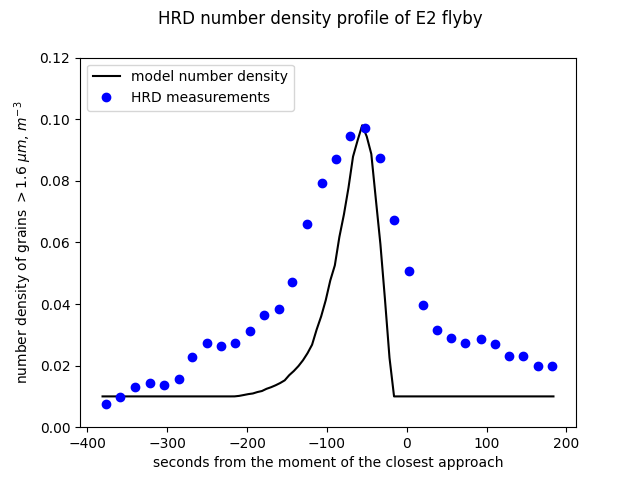}
        \caption{Dust number density profile observed by the HRD during the E2 flyby of the Cassini spacecraft at Enceladus and results from modeling the emission from one single jet in the south polar terrain (see text for details).}
        \label{hrdex}
\end{figure}

On July 14, 2005, the Cassini spacecraft performed a flyby at the Saturnian moon Enceladus (labeled E2). During the flyby, the High Rate Detector (HRD), a subsystem of the Cassini Cosmic Dust Analyzer instrument \citep{Srama:2004uz}, measured the number density of dust particles in the vicinity of the satellite. The significant increase in dust density near Enceladus (see Fig. \ref{hrdex}), about one minute prior to the closest approach of the spacecraft to the satellite, was the first in situ measurement of particles in the Enceladus dust plume \citep{2006Sci...311.1416S}. Dust and vapor are emitted from four fissures called the tiger stripes in the anomalously warm south polar terrain of Enceladus \citep{2006Sci...311.1401S, 2006Sci...311.1393P}.
A part of this dust escapes the moon gravity and forms the dusty E ring of Saturn \citep{2009sfch.book..511H, 2018eims.book..195K}.

From an analysis of high phase-angle images, \citet{2014AJ....148...45P} suggested a list of 100 jets of dust emission for which the coordinates and tilts were derived from images (see also \citet{2015Natur.521...57S}). To demonstrate an application of our model to the dust emission from Enceladus, we selected one single jet from this list with coordinates ($-80.25^\circ$\ N, $ 55.23^\circ$\ E), which is tilted by $5^\circ$ from the surface normal in an azimuthal direction $38^\circ$ away from local north. The ejection is stationary, so that the production rate $\gamma(t)$ is constant. The distributions we implemented for particle sizes, ejection speed, and direction are given by 

\begin{equation}
\label{lognorm}
f_R(R) = \frac{1}{\sigma \sqrt{2\pi}}\frac{1}{R}\exp\left(-\frac{(\ln R - \mu)^2}{2\sigma ^2}\right)
,\end{equation}

\begin{equation}
\label{fu}
f_u(u,R) = \frac{R}{R_c}\left(1 + \frac{R}{R_c}\right)\frac{u}{u_{gas}^2}\left(1 - \frac{u}{u_{gas}}\right)^{\frac{R}{R_c}-1}
,\end{equation}
and
\begin{equation}
\label{unicon}
f_{\tilde{\psi}, \tilde{\lambda_M}}(\tilde{\psi}, \tilde{\lambda}_M)\sin\tilde{\psi} = \left\{ \begin{array}{rl}
\frac{\sin\tilde{\psi}}{1-\cos\omega}\frac{1}{2\pi},\ \tilde{\psi} \le \omega,\\
0,\ \tilde{\psi} > \omega.
\end{array}\right.
\end{equation}
 Equation (\ref{fu}) was derived by \citet{2008Natur.451..685S} to describe the acceleration of dust grains in the gas flux in the vents that supply the sources. Particles smaller than $R_c$ (measured in the same units as $R$) tend to accelerate up to the gas velocity ($u_{gas}$), while particles larger than $R_c$ are significantly slower. For this distribution, we have $u_{min} = 0,$ while $u_{max}$ is equal to the gas velocity $u_{gas}$. Table \ref{enceladus_params} lists the parameters of the distributions and other parameters that are necessary to set up the model.
Figure \ref{hrdex} shows the result for the number density of dust obtained from the model for the single jet, evaluated along the trajectory of Cassini during the E2 flyby. The model was evaluated for grains with a radius larger that 1.6 micron, which corresponds to the size threshold for the HRD data shown in the plot. We multiplied the model profile by a factor so that the peak matches the measured peak density. To match the HRD profile at a large distance from the plume, we added a constant background of 0.01 particles/$m^3$ to the model number density. The selection of the grain size in the model was realized by adjusting the function $G^p_u$ appropriately (see Table \ref{enceladus_params}). We obtained the position of the spacecraft from the reconstructed spice kernels of the mission (https://naif.jpl.nasa.gov/pub/naif/CASSINI/kernels/), using the NAIF Spice toolkit (https://naif.jpl.nasa.gov/naif/toolkit.html). Using these parameters, we recovered the location of the maximum number density on the Cassini trajectory (Fig. \ref{hrdex}). Our model could now be applied to all the jets identified by \citep{2014AJ....148...45P}, and the results could be fit to in situ\emph{} data. Similarly, we could calculate the geometrical optical depths of the dust emitted along a given line of sight and compare this to the brightness distribution in images (see Sect. \ref{ioex} for an example). For a quantitative comparison to images, we can apply light scattering modeling to the dust configuration that is derived from the dust distribution model.
\begin{table*}
        \caption{Parameters used to model the number density profile of the E2 flyby}
        \begin{tabular}{|c|c|}
                \hline
                Parameter & Comment \\
                \hline
        $\gamma(t) = 1.35 \cdot 10^{14} \ s^{-1}$ & The dust ejection is stationary, the rate \\
        & was chosen to fit the data\\
        \hline
        Eq. (\ref{lognorm}) with $\mu = -1.0$ and $\sigma = 1.5$  & Small particles dominate the population,\\
        is used as a size distribution & but the distribution is not too steep\\
        \hline
         Eq. (\ref{fu}) with $R_c=0.5\ \mu m$  & This expression describes  \\
         and $u_{gas} = 1000$ m/s is used  & the dust acceleration by the gas flux \\
         as the ejection speed distribution&inside the channels as it is at Enceladus\\
         \hline
         Eq. (\ref{unicon}) with $\omega=10^\circ$ is used& The distribution describes uniform emission \\
         as the distribution  & into a cone of a given width\\
         of the ejection direction&\\
         \hline
         $R_{min} = 1.6 \ \mu m$ & The lower sensitivity threshold of the HRD \\
         \hline
          & An arbitrarily chosen but reasonably  \\
           &high value, meaning that we can neglect\\
           & particles with larger sizes.  With our\\
         $R_{max} = 6 \ \mu m$ & choice of size distribution and the size  \\
         & -dependent ejection speed distribution, it \\
          &is highly unlikely that the HRD detects \\
          & a particle of this size at the altitude\\
          &  of E2 flyby\\
          \hline
         $p=0$ & We are interested in the number density \\
         \hline
        \end{tabular}
        \label{enceladus_params}
\end{table*}

\subsection{Surface deposition of material from plumes on Europa}
\label{europaex}
There is evidence that cryovolcanic activity also generates plumes on the Jupiter satellite Europa \citep{2000Icar..144...54F, 2000JGR...10522579P, 2014Sci...343..171R,Sparks:2016gl,Jia:2018ii}. Owing to the higher gravity of Europa, plumes would be more confined than those on Enceladus, and they may be harder to observe or sample directly \citep{2015GeoRL..4210541S, Quick:2020kp}. However, surface features from plume deposits may provide evidence for past cryovolcanic activity, and if spectral features exist, allow the remote characterization of material from the interior \citep{2000Icar..144...54F,2000JGR...10522579P,Quick:2020kp}.

We calculated the radial variation of the mass flux of dust grains falling back onto the surface, considering four different dust sources with identical characteristics except for the particle size distribution and the distribution of ejection directions. We considered two size distributions and two ejection modes: one describing a narrow jet, and the other describing a broader, more diffuse ejection. For the size distributions we employed a power law

\begin{equation}
\label{pow}
f_R(R) = \frac{1-q}{R_2^{1-q}-R_1^{1-q}}R^{-q}
,\end{equation}
with two different values of the exponent $q$. The greater $q$, the more abundant the small dust particles. For the ejection direction, we used a pseudo-Gaussian distribution of the polar angle 
\begin{equation}
\label{Gauss}
f_{\tilde{\psi}, \tilde{\lambda}_M}(\tilde{\psi}, \tilde{\lambda}_M)\sin\tilde{\psi} = C_{norm}e^{-(\tilde{\psi}-\psi_{max})^2/2\omega^2} \frac{\sin\tilde{\psi}}{2\pi}
,\end{equation}
giving a nonzero probability of ejection in any direction. The normalization constant $C_{norm}$ was found numerically for fixed values of $\psi_{max}$ and $\omega$. The dust production rate $\gamma$ was identical for all the sources, meaning that they produced the same number of dust particles per unit time, but we then obtained different mass production rates for the different size distributions. For the ejection speed distribution we again used Eq. (\ref{fu}). Table \ref{europa_params} lists the distribution parameters.

We can compute the rate of dust mass produced per second from the size distribution as
\begin{equation}
        \label{totmass}
        \frac{\mathrm{d}m}{\mathrm{d}t} =\gamma\rho\frac{4\pi}{3}\int_{R_{min}}^{R_{max}}f_R(R)R^3\mathrm{d}R,
\end{equation}
where $\rho$ is the density of the ice grains, and the grains were assumed to be spherical.

To compute flux instead of density, we must modify Eq. (\ref{workformula}) as
\begin{multline}
        \label{flux}
        n(r,\alpha,\beta,R_{min}<R<R_{max}) = \frac{\gamma}{rr_M\sin\Delta\phi} \int_{v_{min}}^{v_{max}}\frac{G^p_u(R_{min},R_{max})}{u^2}\times\\\times \sum_{i}\frac{f_{\psi,\lambda_M}(\psi_i,\lambda_{Mi})\sin\psi_i}{\cos\psi_i}\abs*{\frac{\partial\Delta\phi}{\partial\theta}}_{\theta^*_i}^{-1}v^2|\cos\theta_i^*|dv.  
\end{multline}
As  the vector $\mathbf{r}$ is normal to the surface of the satellite, the factor $v|\cos\theta|$ turns the density into the flux of dust falling back to the moon. Multiplying equation \ref{flux} by $4\pi\rho/3 $ ($\rho$ is the density of the particle material) and setting $p=3$ and $r=r_M$ gives the mass flux onto the surface. Because $r = r_M$ on the surface, it is geometrically impossible to obtain $\theta^*_i < \pi/2$ or any particles moving upward.

\begin{figure}[H]
        \centering
        \includegraphics[width=0.5\textwidth]{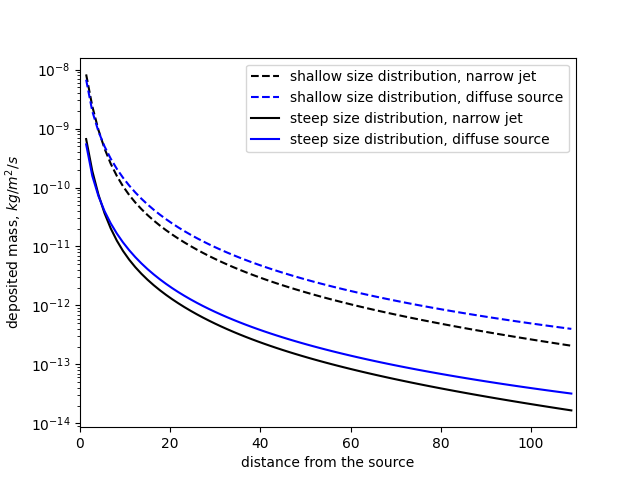}
        \caption{Dust flux on the surface of the moon at various distances from the dust source. The production rate of the sources with shallow dust size distribution is 0.6 kg/s, and the production rate of the sources with steep size distribution is 0.05 kg/s.}
        \label{deposits}
 \end{figure}

With the parameters from Table \ref{europa_params}, Eq (\ref{totmass}) gives $0.6$ kg of dust produced each second for the source with the shallow size distribution ($q=3$) and $0.05$ kg for the source with the steep size distribution ($q=5$). Figure \ref{deposits} shows the distribution of the dust deposition (mass flux onto the surface) with distance from the source on the moon surface.
 \begin{table*}
        \caption{Parameters used to model the radial distribution of dust deposition on the surface of Europa}
        \begin{tabular}{|c|c|}
                \hline
                Parameter & Comment \\
                \hline
        $\gamma(t) = 10^{14} \ s^{-1}$ & The dust ejection is considered stationary\\
                \hline
        Eq.( \ref{pow}) with $q = 3$, $R_1 = 0.2\ \mu m$& We paired these size and ejection \\
          and $R_2 = 20\ \mu m$ as   & direction distributions\\
           the shallow\ size distribution & to model four sources: a narrow jet\\
         and with the same boundaries, but $q=5$ & with a shallow size distribution,\\
         as the steep\ size distribution & a narrow jet with a steep size distribution,\\
         & a diffuse source  with a shallow size\\
         Eq. (\ref{Gauss}) with $\omega=5^\circ$ and $\psi_{max} = 0^\circ$&  distribution, and a diffuse source\\
         for the  narrow jet\ and with $\omega=45^\circ$ & with a steep size distribution \\
         and $\psi_{max}=45^\circ$ for the  diffuse source& \\
        \hline
                Eq. (\ref{fu}) with $R_c=0.5\ \mu m$  & This expression describes  \\
                and $u_{gas} = 700$ m/s is used&the acceleration of the dust by the gas  \\
                 as the ejection speed distribution &flux inside the vents as it is thought be at Europa \\
                \hline
 & We wish to compute the total mass\\
                $R_{min} = 0.2 \ \mu m$ &  produced by each source, \\
                $R_{max} = 20 \ \mu m$ & therefore the interval $(R_{min},R_{max})$ \\
 &coincides with $(R_1,R_2)$\\
                \hline
                $p=3$ & We are interested in the mass \\
                \hline
                $\rho = 920\ kg/m^3$ & Water-ice density\\
                \hline
        \end{tabular}
        \label{europa_params}
 \end{table*}

Taking the observational evidence together, we expect the outgassing activity on Europa to be intermittent or episodic, with a potentially complex time dependence. This case could be modeled using a time-dependent function for the dust production rate. We consider a time-dependent dust ejection in our final example.
\subsection{Images of a volcanic eruption on Io}
\label{ioex}
The innermost satellite of Jupiter, Io, is a geologically very active body. It has multiple volcanic centers scattered over its surface \citep{1979Natur.280..733S,2001JGR...10633025K, 2004Icar..169...29G}. The volcanic plumes vary in many features as shape, size and period of activity and often they are asymmetric. Our model allows to investigate such plumes by using inclined jets with time-dependent eruptions, or even a simultaneous dust ejection from sources with different properties.

The main flaw of our model in application to the Io volcanic plumes is that the gas emerging from the vents is generally not collisionless. In the dense cores of the plumes, condensation occurs up to a height of several hundred kilometers \citep{2002JGRE..107.5109C}. Moreover, the dynamics of the dust will be influenced by the surrounding gas. Condensation (i.e.,\ dust production) in the plume might be taken into account in principle by modeling such a plume with additional point sources placed within the condensation column, with appropriately adjusted time-dependent dust production rates. For this purpose, we implemented in the code the possibility of computing the dust density also at locations below (closer to the moon center) the location of the source. In this case, we still use the assumption that the true anomaly at the location of the source is lower than $\pi$, that is, downward ejection from a source located above the surface is not possible. 

In our simple example presented here, we considered just one source located on the surface to represent a volcanic plume with a finite duration of activity. We took several images of the expanding dust plume from the same point in space with a favorable geometry when the volcano is located at the limb of the moon. We also assumed that the glow of the surface of Io was already subtracted from the images but a homogeneous background brightness was present, so that the moon disk looks dark against this background. We tilted the volcano slightly by $3^\circ$ toward local south. The volcano was not exactly in the center of the image for the purpose of reducing the computational difficulties arising when $\Delta\phi$ or $\Delta\beta \approx 0$ (see Sect. \ref{spepoints}). 

Synthetic images of size 128x128 pixels were then constructed in the following way. Each pixel corresponds to a line of sight. We placed a grid of points on the line of sight for which we calculated the total cross section covered by the dust grains ($p=2$ in Eq. (\ref{Gu})). Then we integrated over the lines of sight to obtain the geometrical optical depth (total particle cross section per area covered by the pixel). The color of a pixel then corresponded to the value determined for the geometric optical depth.

We employed Eq. (\ref{pow}) for the size distribution, Eq. (\ref{Gauss}) for the distribution of the ejection direction, and a simple size-independent expression (Eq. (\ref{fuun})) for a uniform ejection speed distribution,
\begin{equation}
\label{fuun}
f_u(u,R) = \frac{1}{u_{max} - u_{min}}
.\end{equation}

\begin{figure*}[h!]
        \centering
        \begin{minipage}{0.33\textwidth}
                \centering
                \includegraphics[width=0.95\textwidth]{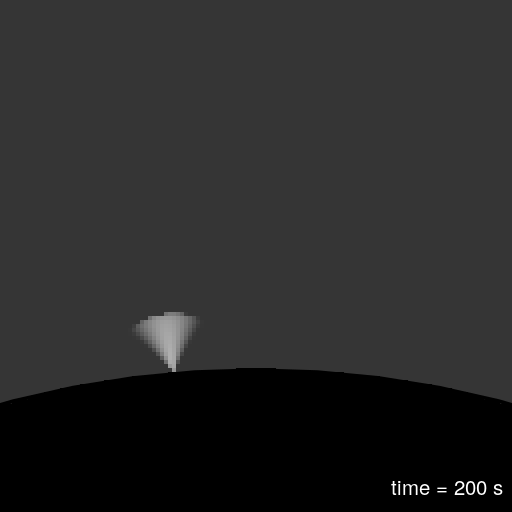} 
        \end{minipage}\hfill
        \begin{minipage}{0.33\textwidth}
                \centering
                \includegraphics[width=0.95\textwidth]{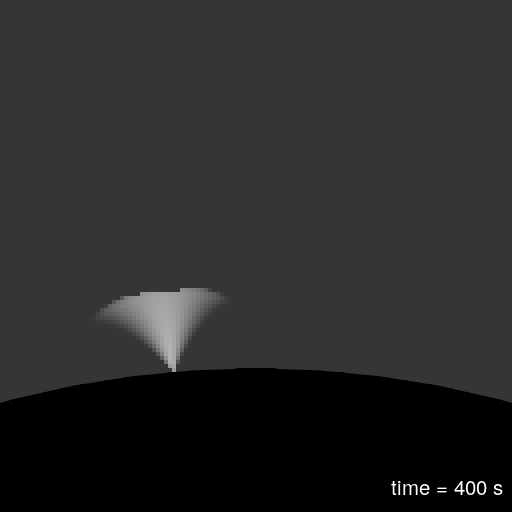} 
        \end{minipage}\hfill
        \begin{minipage}{0.33\textwidth}
                \centering
                \includegraphics[width=0.95\textwidth]{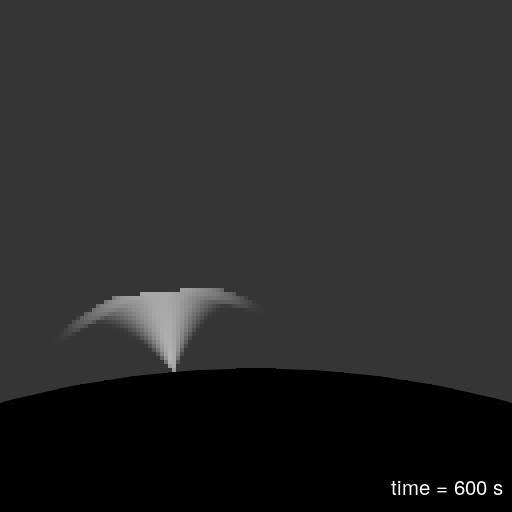} 
        \end{minipage}\hfill\\
        \vskip0.1cm
        \begin{minipage}{0.33\textwidth}
                \centering
                \includegraphics[width=0.95\textwidth]{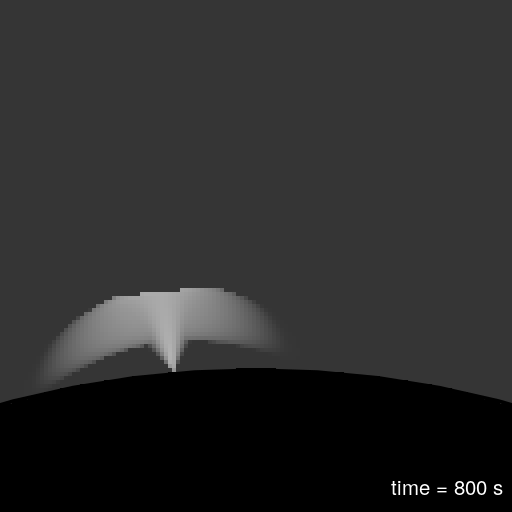} 
        \end{minipage}\hfill
        \begin{minipage}{0.33\textwidth}
                \centering
                \includegraphics[width=0.95\textwidth]{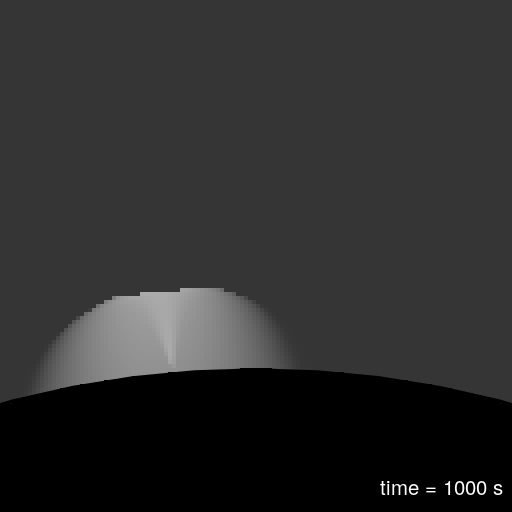} 
        \end{minipage}\hfill
        \begin{minipage}{0.33\textwidth}
                \centering
                \includegraphics[width=0.95\textwidth]{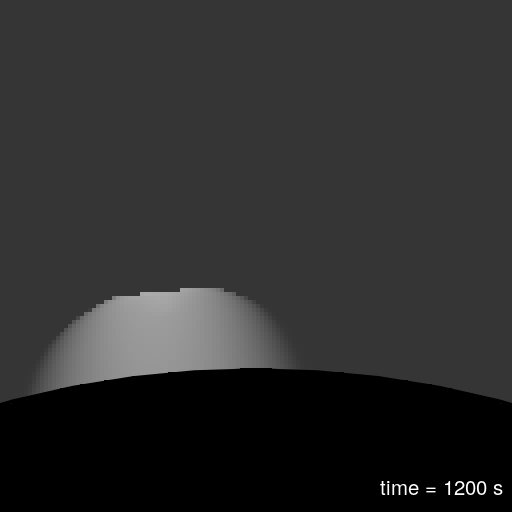} 
        \end{minipage}\hfill
        \vskip0.1cm
        \begin{minipage}{0.33\textwidth}
                \centering
                \includegraphics[width=0.95\textwidth]{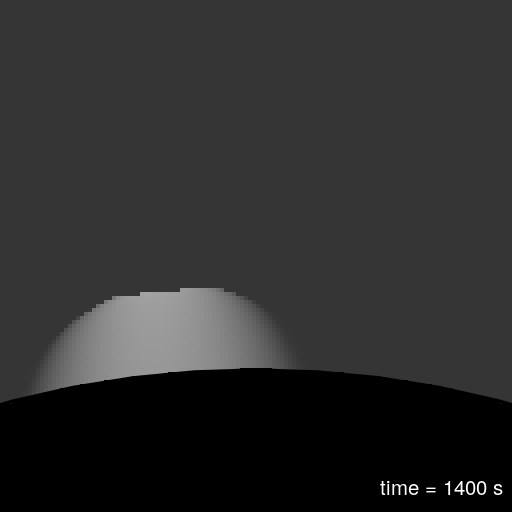} 
        \end{minipage}\hfill
        \begin{minipage}{0.33\textwidth}
                \centering
                \includegraphics[width=0.95\textwidth]{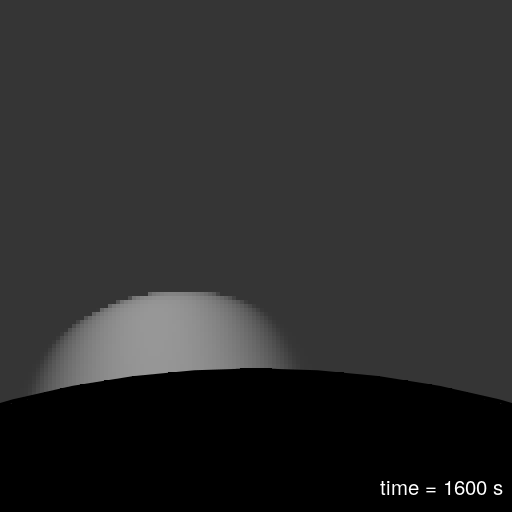} 
        \end{minipage}\hfill
        \begin{minipage}{0.33\textwidth}
                \centering
                \includegraphics[width=0.95\textwidth]{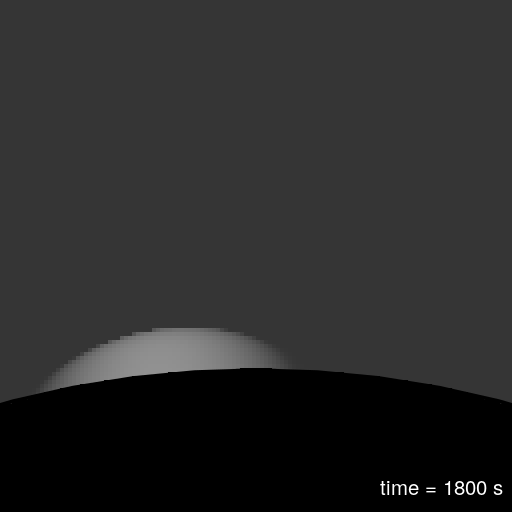} 
        \end{minipage}\hfill
        \caption{Images of a fictive volcanic plume on Io taken at different stages of eruption.}
        \label{9im}
\end{figure*}

When the time interval in which the plume is observed is much shorter than the characteristic time of the plume variability, the ejection can be considered steady. However, here we modeled a nonstationary volcanic plume that was active for only 1000 seconds and constructed nine snapshots of this plume to show the evolution of the space distribution of dust with time 
\begin{equation}
\label{gammat}
\gamma(t) = \left\{ \begin{array}{rl}
\gamma_0\frac{-t^2+2t_{max}t}{t_{max}^2} ,\ 0 < t < 2t_{max},\\
0,\ t < 0\ \text{or}\ t > 2t_{max},
\end{array}\right
.\end{equation}
where $t_{max} = 500$ s. Table \ref{io_params} describes the model setup, and the images are shown in Fig. \ref{9im}
\begin{table*}
        \caption{Parameters used to construct the volcanic plume images.}
        \begin{tabular}{|c|c|}
                \hline
                Parameter & Comment \\
                \hline
                Eq. (\ref{gammat}) with $t_{max} = 500$ s  & The volcanic plume \\
                and $\gamma_0 = 10^{14} \ s^{-1}$ as the time- & ejected dust for 1000 s\\
                dependent dust production rate& \\
                \hline
                Eq. (\ref{pow}) with $q = 3.0$, & \\
                 $R_1 = 0.2\ \mu m$, and $R_2 = 20\ \mu m$  & \\
                is used as the size distribution & \\
                \hline
                The ejection speed is uniformly  & The narrow range of initial  \\
                distributed between &velocities allows us to obtain \\
                700 m/s and 750 m/s & the umbrella-shaped plume\\
                \hline
                Eq. (\ref{Gauss}) with $\psi_{max} = 0^\circ$ & The jets are very narrow,   \\
                and $\omega=5^\circ$ is used as the distribution  &but there is a nonzero probability \\
                of the ejection direction & of ejection in any direction\\
                \hline
                $R_{min} = 0.2 \ \mu m$ & We observe the particles with the sizes \\
                $R_{max} = 0.4 \ \mu m$ &  close to the optical wavelength range, \\
                 & so that the radii are twice smaller\\
                \hline
                $p=2$ & We are interested in the area \\
                & covered by the particles \\
                \hline
        \end{tabular}
        \label{io_params}
\end{table*}

Compared to the examples from Sects \ref{enceladusex} and \ref{europaex}, the construction of images is computationally expensive. Therefore it makes sense to avoid calculations of the properties in those points where the value would not affect the final result. In the case considered here, the maximum ejection velocity was set to be lower than the escape velocity of Io, so that there is a maximum height that the particles can reach. Moreover, the disk of the moon covers part of the images. Taking these two facts into account, we excluded the points on the lines of sight from the calculations that crossed the moon disk and the points that lie at a greater distance from the surface than the maximum height.

\section{Discussion}
We developed a semianalytical model that allows the user to derive the average properties of dust (number or mass density, fluxes, and optical depths) ejected from an atmosphereless body. Physically, our approach is based on the two-body problem, that is,\ it neglects any forces on the dust particles other than the point-mass gravity of the source body. We show that the model still has a wide range of applications. These are situations where the gravity of other perturbing bodies and higher-order gravity from a nonspherical mass distribution of the central body can be neglected along the entire path that a dust particle takes from its source to its sink. The nongravitational forces must also be negligibly small, for instance, electromagnetic forces acting on charged grains, radiation-induced forces (solar radiation pressure and Poynting-Robertson drag), and drag exerted by ambient gas or plasma. For instance, the model can be applied to the Enceladus dust plume in a region that is sufficiently close to the dust sources on the south polar terrain of this satellite. It cannot be applied to estimate the dust density at higher northern latitudes of Enceladus, however, because this region is too far away from the sources near the south pole, and three-body forces due to Saturn have already affected the shape of the dust configuration. This can be seen in a peculiar pattern of color variation \citep{2011Icar..211..740S}, that can be matched with two wedge shaped regions extending deep into the northern hemisphere \citep{2011epsc.conf.1358S,2017LPI....48.2601S}, for which three body models of the plume predict enhanced fall back rates of south polar dust \citep{2010Icar..206..446K}. The model can be applied to investigate the dust clouds around the Galilean moons \citep{1999Natur.399..558K} at any longitude and latitude, however, because in this case, the dust emission occurs (nearly) uniformly over the whole surface of the satellites as long as the point of interest is located deeply enough in the Hill sphere of the satellite. 

Mathematically, our model relates the dust distribution at the site of ejection to the distribution at the point of interest (spacecraft position) by using the conservation laws of energy and momentum provided by the two-body problem \citep{2003P&SS...51..251K, Sremcevic:2003gf}. We use the fact that the position of the spacecraft, the position of the source, and the center of the moon define the plane to which the movement is restricted. In the evaluation of the dust properties at a given point, this allows us to carry out two of the three integrations over velocity space analytically. Only one remaining integration must then be performed numerically. The distribution-based approach is very flexible, and because it allows employing asymmetric and nonstationary modes of dust emission, it allows modeling quite complex situations. A time dependence could relatively easily be introduced in the distributions of ejection speed, direction, and size as well. Furthermore, the model could also be extended to model dust emission from a comet or an active asteroid, making the Sun the central body and using a rescaled solar mass to account for the radiation pressure.

Relying only on one numerical integration, the model becomes computationally very efficient, so that even image reconstruction becomes feasible, although it involves the evaluation of the dust properties in the three-dimensional region that is in the field of view. From the examples discussed in Sect. \ref{sec:applications}, the image of the volcano (Sect. \ref{ioex}) is the most computationally demanding. Nevertheless, on a usual four-core PC, each image in Fig. \ref{9im} required 0.4 s of elapsed time to be obtained.

 The model is implemented in Fortran-95, and the package called DUDI is available at \\
 https://github.com/Veyza/dudi for free usage under the terms of GNU General Public License. A user can use the  probability density functions and choose from several variants that are already implemented in the described examples, or they can implement new variants.  

\section*{Acknowledgements}
This work was supported by the Academy of Finland.

\section*{Appendix A. Replacement of the variable in the argument of Dirac's $\delta$-function}
\label{replsec}

Let $x\in S \subset \mathbb{R}^n$; $f,g : S \rightarrow \mathbb{R}^n$. Then we can perform a replacement of the variable under the integral based on Eqs. (\ref{dg}) and (\ref{dint}) from \citet{GelfShil},
\begin{equation}
\label{dg}
\delta(g(x)) = \sum_{i}\frac{\delta(x-x_i)}{|g'(x_i)|}
.\end{equation}
Here, $g'$ is the Jacobian matrix of the function $g$, $|g'|$ denotes the Jacobi determinant, and $x_i$ are the zeros of $g$,
\begin{equation}
\label{dint}
\int_{\mathbb{R}^n}\delta(g(x))f(g(x))|g'(x)|dx = \int_{g(S)}\delta(h)f(h)dh
\end{equation}

\begin{eqnarray}
\nonumber
\int_S \delta(g(x))f(x)dx &=& \int_S \delta(g(x))f(g^{-1}(g(x)))dx\\
\nonumber
&=& \int_S\delta(g(x))f^*(g(x))dx \\
\nonumber
&=&\int_S \delta(g(x))\frac{f^*(g(x))}{| g'(x)|}|g'(x)|dx \\
\nonumber
&=&\int_S \delta(g(x))\frac{f^*(g(x))}{| g'(g^{-1}(g(x)))|}|g'(x)|dx \\
\nonumber
&=&\int_S \delta(g(x))f^*_D(g(x))|g'(x)|dx \\
\nonumber
&=& \int_{g(S)}\delta(h)f^*_D(h)dh,
\end{eqnarray}
where
\begin{equation*}
f^*_D(h) = \frac{f(x)}{\abs*{g'(x)}},\quad h = g(x).
\end{equation*}
Upon integration, we obtain
\begin{equation*}
\int_{g(S)}\delta(h)f^*_D(h)dh = \sum_{i}\frac{f(x_i)}{\abs*{g'(x_i)}},\ \ g(x_i) = \textbf{0}.
\end{equation*}

In our case, $n=2,$ the function $g$ performs a transformation from $(\alpha_M,\beta_M)$ to $(\theta,\lambda),$ and the function $f$ represents all the dependences on $\theta$ and $\lambda$ in the integrand in Eq. (\ref{ndens}).

\section*{Appendix B. Physical meaning of the function $G_u$}
For each source, the first step in the numerical calculations is to obtain values of $G^p_u(R_{min}, R_{max})$ (Eq. (\ref{Gu})) on a dense grid of $u$. This means that we perform integrations over the particle size $R$ treating $u$ as a parameter. The size distribution $f_R(R)$ can be defined on an interval of grain sizes, its lower and upper boundaries being parameters of the distribution. 

\begin{figure}
        \centering
        \begin{minipage}{0.45\textwidth}
                \centering
                \includegraphics[width=0.9\textwidth]{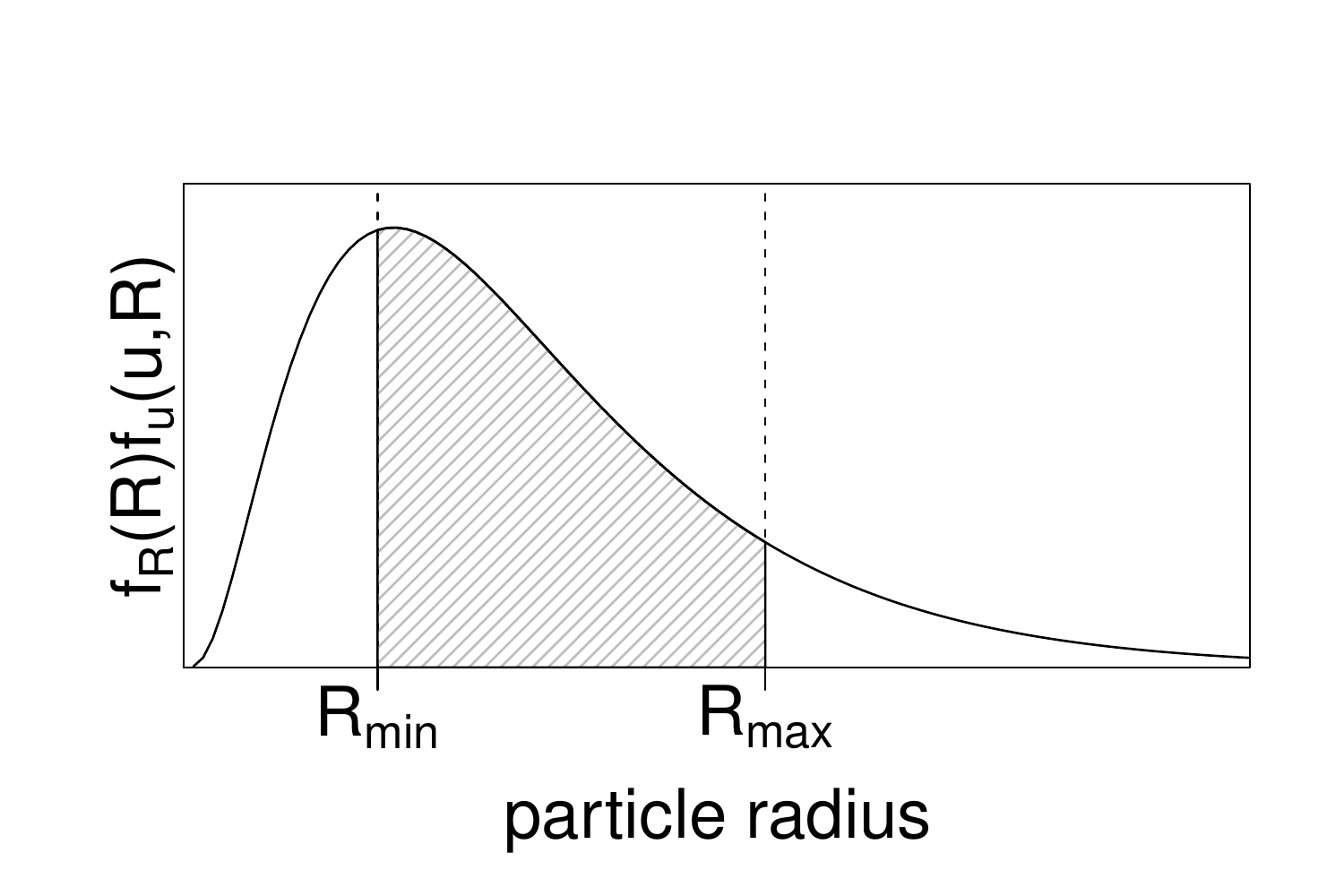} 
                \caption{Integrand in Eq (\ref{Gu}) with a lognormal size distribution and a fixed value of $u.$}
                \label{GRexlognorm}
        \end{minipage}\hfill
        \begin{minipage}{0.45\textwidth}
                \centering
                \includegraphics[width=0.9\textwidth]{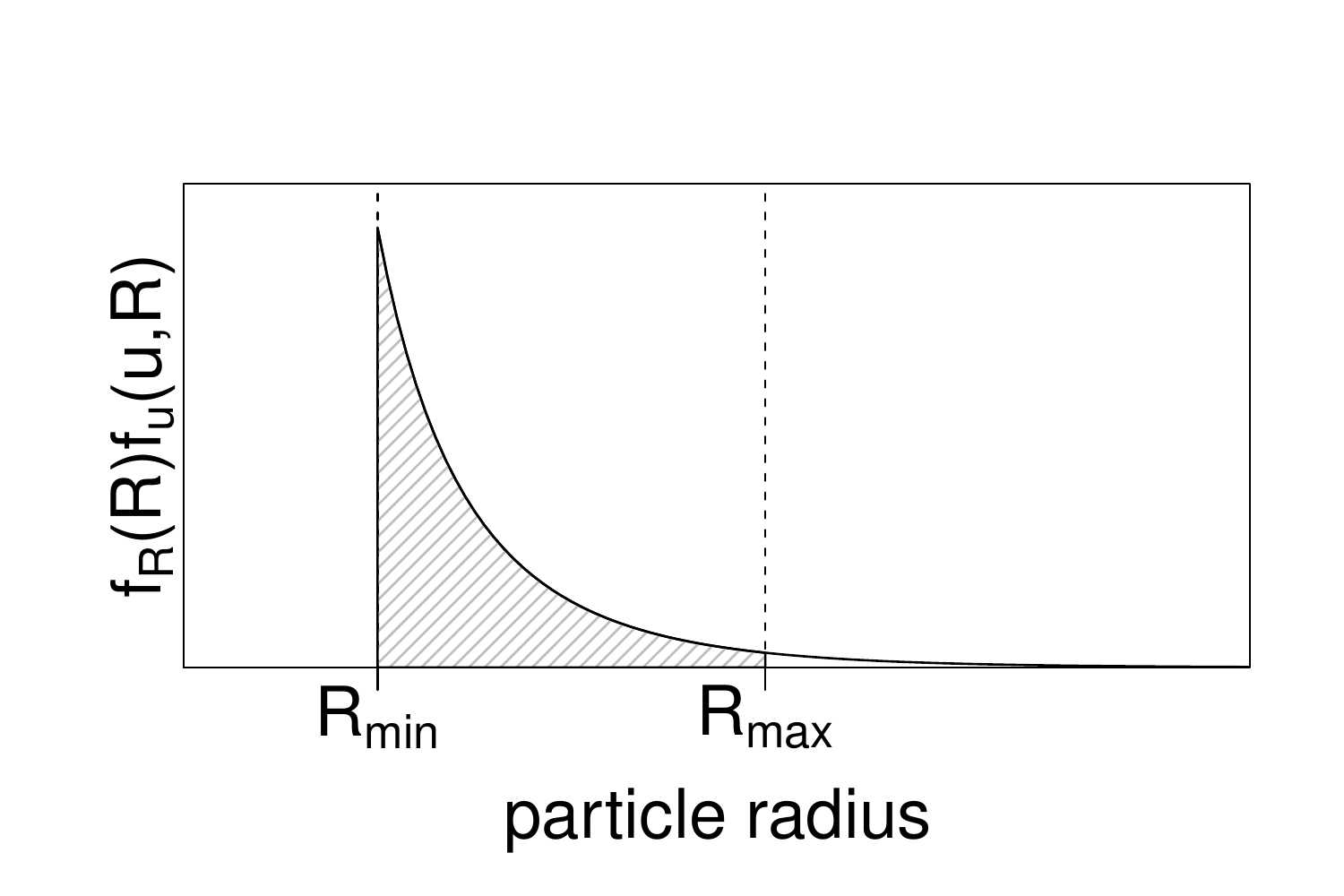} 
                \caption{Integrand in Eq. (\ref{Gu}) with a power-law size distribution and a fixed value of $u.$}
                \label{GRexpower}
        \end{minipage}\hfill
        \captionsetup{labelformat=empty}
\end{figure}

These boundaries need not necessarily be equal to $R_{min}$ and $R_{max}$. Instead, $R_{min}$ and $R_{max}$ should be understood as limits of an observable interval of particle radii. Particles smaller than $R_{min}$ or larger than $R_{max}$ may exist, and they contribute to the normalization of $f_R(R)$, but they would not contribute to the value of $n$ (Eq. (\ref{workformula})). This concept is illustrated in Figs. \ref{GRexlognorm} and \ref{GRexpower}. The quantity plotted in the two figures is $f_R(R)f_u(u,R)$. In Fig. \ref{GRexlognorm} ,$f_R(R)$ is a lognormal distribution that is formally defined in the interval $(0,+\infty)$, or in other words, $\int_{0}^{+\infty}f_R(R)dR = 1$, and the interval $(R_{min}, R_{max})$ does not cover the whole domain of $f_R(R)$. In Fig. \ref{GRexpower}, the particle sizes are distributed between certain $R_1$ and $R_2$ as Eq. (\ref{pow}). In this example, $R_{min} = R_1$ , but $R_{max} < R_2$. If $R_{min}$ were lower than $R_1$ , it would not have changed the result because for $R < R_1$ , we have $f_R(R) = 0$.

Although we defined the size distribution outside the interval $(R_{min}, R_{max}),$ its shape there does not affect the final result as long as it remains the same inside $(R_{min}, R_{max})$. However, we can investigate the defined size distribution by applying different values of $R_{min}$ and $R_{max}$. This approach implies that the size distribution and the initial speed distribution are physical characteristics of the dust source, while $R_{min}$ and $R_{max}$ represent the sensitivity range of the instrument with which we perform observations. In the model implementation DUDI, the interval $(R_{min},R_{max})$ can coincide with the $f_R(R)$ domain or can be even larger.

The units of the particle radius $R$ matter only in the definition of $f_R$ and $f_u$, so that we suggest measuring $R$ in microns to have simpler numbers in the expressions.
With different formulae for $f_R$ and $f_u$ , shorter expressions for $G^p_u$ can be obtained that require less cumbersome computations (see, e.g., \citet{2011Natur.474..620P}). However, we purposefully consider a set of $f_R$ and $f_u$ in our model that cannot be simplified to show the general form of the solution and to allow flexibility in applying the model.
\newpage
\bibliographystyle{apalike}
\bibliography{PapersLit,AdditionalLib}

\begin{thebibliography}{}

\bibitem[Cataldo et~al., 2002]{2002JGRE..107.5109C}
Cataldo, E., Wilson, L., Lane, S., and Gilbert, J. (2002).
\newblock {A model for large-scale volcanic plumes on Io: Implications for
  eruption rates and interactions between magmas and near-surface volatiles}.
\newblock {\em Journal of Geophysical Research}, 107(E11, 5109).

\bibitem[Fagents et~al., 2000]{2000Icar..144...54F}
Fagents, S.~A., Greeley, R., Sullivan, R.~J., Pappalardo, R.~T., Prockter,
  L.~M., and Team, T. G.~S. (2000).
\newblock {Cryomagmatic Mechanisms for the Formation of Rhadamanthys Linea,
  Triple Band Margins, and Other Low-Albedo Features on Europa}.
\newblock {\em Icarus}, 144(1):54--88.

\bibitem[Geissler et~al., 2004]{2004Icar..169...29G}
Geissler, P., McEwen, A., Phillips, C., Keszthelyi, L., and Spencer, J.~R.
  (2004).
\newblock {Surface changes on Io during the Galileo mission}.
\newblock {\em Icarus}, 169(1):29--64.

\bibitem[Gelfand and Shilov, 1968]{GelfShil}
Gelfand, I.~M. and Shilov, G.~E. (1968).
\newblock {\em {Generalized functions}}, volume~1.
\newblock Academic press.

\bibitem[{Hor{\'a}nyi} et~al., 2009]{2009sfch.book..511H}
{Hor{\'a}nyi}, M., {Burns}, J.~A., {Hedman}, M.~M., {Jones}, G.~H., and
  {Kempf}, S. (2009).
\newblock {\em {Diffuse Rings}}, page 511.

\bibitem[Hor{\'a}nyi et~al., 2015]{2015Natur.522..324H}
Hor{\'a}nyi, M., Szalay, J.~R., Kempf, S., Schmidt, J., Gr{\"u}n, E., Srama,
  R., and Sternovsky, Z. (2015).
\newblock {A permanent, asymmetric dust cloud around the Moon}.
\newblock {\em Nature}, 522(7556):324--326.

\bibitem[Jia et~al., 2018]{Jia:2018ii}
Jia, X., Kivelson, M.~G., Khurana, K.~K., and Kurth, W.~S. (2018).
\newblock {Evidence of a plume on Europa from Galileo magnetic and plasma wave
  signatures}.
\newblock {\em Nature Astronomy}, 395:1.

\bibitem[Kempf et~al., 2010]{2010Icar..206..446K}
Kempf, S., Beckmann, U., and Schmidt, J. (2010).
\newblock {How the Enceladus dust plume feeds Saturn{\textquoteright}s E ring}.
\newblock {\em Icarus}, 206(2):446--457.

\bibitem[{Kempf} et~al., 2018]{2018eims.book..195K}
{Kempf}, S., {Hor{\'a}nyi}, M., {Hsu}, H.~W., {Hill}, T.~W., {Juh{\'a}sz}, A.,
  and {Smith}, H.~T. (2018).
\newblock {\em {Saturn's Diffuse E Ring and Its Connection with Enceladus}},
  page 195.

\bibitem[Keszthelyi et~al., 2001]{2001JGR...10633025K}
Keszthelyi, L., McEwen, A., Phillips, C., Milazzo, M., Geissler, P., Turtle,
  E., Rade-baugh, J., Williams, D., Simonelli, D., Breneman, H., Klaasen, K.,
  Levanas, G., Denk, T., and Team, G.~S. (2001).
\newblock {Imaging of volcanic activity on Jupiter’s moon Io by Galileo
  during the Galileo Europa Mission and the Galileo Millennium Mission}.
\newblock {\em J. Geophys. Res}, 106:33025–33052.

\bibitem[Krivov et~al., 2003]{2003P&SS...51..251K}
Krivov, A.~V., Srem{\v c}evi{\'c}, M., Spahn, F., Dikarev, V., and
  Kholshevnikov, K.~V. (2003).
\newblock {Impact-generated dust clouds around planetary satellites:
  spherically symmetric case}.
\newblock {\em Planetary and Space Science}, 51(3):251--269.

\bibitem[Kr{\"u}ger et~al., 1999]{1999Natur.399..558K}
Kr{\"u}ger, H., Krivov, A.~V., Hamilton, D.~P., and Gr{\"u}n, E. (1999).
\newblock {Detection of an impact-generated dust cloud around Ganymede}.
\newblock {\em Nature}, 399(6):558--560.

\bibitem[Kr{\"u}ger et~al., 2003]{2003Icar..164..170K}
Kr{\"u}ger, H., Krivov, A.~V., Srem{\v c}evi{\'c}, M., and Gr{\"u}n, E. (2003).
\newblock {Impact-generated dust clouds surrounding the Galilean moons}.
\newblock {\em Icarus}, 164(1):170--187.

\bibitem[Phillips et~al., 2000]{2000JGR...10522579P}
Phillips, C.~B., McEwen, A.~S., Hoppa, G.~V., Fagents, S.~A., Greeley, R.,
  Klemaszewski, J.~E., Pappalardo, R.~T., Klaasen, K.~P., and Breneman, H.~H.
  (2000).
\newblock {The search for current geologic activity on Europa}.
\newblock {\em Journal of Geophysical Research}, 105(E):22579--22598.

\bibitem[Porco et~al., 2014]{2014AJ....148...45P}
Porco, C., DiNino, D., and Nimmo, F. (2014).
\newblock {How the Geysers, Tidal Stresses, and Thermal Emission across the
  South Polar Terrain of Enceladus are Related}.
\newblock {\em The Astronomical Journal}, 148(3):45.

\bibitem[Porco et~al., 2006]{2006Sci...311.1393P}
Porco, C.~C., Helfenstein, P., Thomas, P.~C., Ingersoll, A.~P., Wisdom, J.,
  West, R.~A., Neukum, G., Denk, T., Wagner, R., Roatsch, T., Kieffer, S.,
  Turtle, E.~P., McEwen, A., Johnson, T.~V., Rathbun, J., Veverka, J., Wilson,
  D., Perry, J., Spitale, J.~N., Brahic, A., Burns, J.~A., Del~Genio, A.~D.,
  Dones, L., Murray, C.~D., and Squyres, S. (2006).
\newblock {Cassini Observes the Active South Pole of Enceladus}.
\newblock {\em Science}, 311(5):1393--1401.

\bibitem[Postberg et~al., 2011]{2011Natur.474..620P}
Postberg, F., Schmidt, J., Hillier, J.~K., Kempf, S., and Srama, R. (2011).
\newblock {A salt-water reservoir as the source of a compositionally stratified
  plume on Enceladus}.
\newblock {\em Nature}, 474(7):620--622.

\bibitem[Quick and Hedman, 2020]{Quick:2020kp}
Quick, L.~C. and Hedman, M.~M. (2020).
\newblock {Characterizing deposits emplaced by cryovolcanic plumes on Europa}.
\newblock {\em Icarus}, 343:113667--15.

\bibitem[Roth et~al., 2014]{2014Sci...343..171R}
Roth, L., Saur, J., Retherford, K.~D., Strobel, D.~F., Feldman, P.~D., McGrath,
  M.~A., and Nimmo, F. (2014).
\newblock {Transient Water Vapor at Europa's South Pole}.
\newblock {\em Science}, 343:171--174.

\bibitem[{Schenk} et~al., 2017]{2017LPI....48.2601S}
{Schenk}, P., {Buratti}, B., {Helfenstein}, P., {Kempf}, S., and {Schmidt}, J.
  (2017).
\newblock {Colors of Enceladus: Plume Redeposition and Lessons for Europa}.
\newblock In {\em Lunar and Planetary Science Conference}, Lunar and Planetary
  Science Conference, page 2601.

\bibitem[Schenk et~al., 2011a]{2011Icar..211..740S}
Schenk, P., Hamilton, D.~P., Johnson, R.~E., McKinnon, W.~B., Paranicas, C.,
  Schmidt, J., and Showalter, M. (2011a).
\newblock {Plasma, plumes and rings: Saturn system dynamics as recorded in
  global color patterns on its midsize icy satellites}.
\newblock {\em Icarus}, 211(1):740--757.

\bibitem[Schenk et~al., 2011b]{2011epsc.conf.1358S}
Schenk, P., Schmidt, J., and White, O. (2011b).
\newblock {The Snows of Enceladus}.
\newblock {\em EPSC-DPS Joint Meeting 2011}, page 1358.

\bibitem[Schmidt et~al., 2008]{2008Natur.451..685S}
Schmidt, J., Brilliantov, N.~V., Spahn, F., and Kempf, S. (2008).
\newblock {Slow dust in Enceladus' plume from condensation and wall collisions
  in tiger stripe fractures}.
\newblock {\em Nature}, 451(7):685--688.

\bibitem[Southworth et~al., 2015]{2015GeoRL..4210541S}
Southworth, B.~S., Kempf, S., and Schmidt, J. (2015).
\newblock {Modeling Europa's dust plumes}.
\newblock {\em Geophysical Research Letters}, 42(2):10--.

\bibitem[Spahn et~al., 2006]{2006Sci...311.1416S}
Spahn, F., Schmidt, J., Albers, N., Horning, M., Makuch, M., Sei{\ss}, M.,
  Kempf, S., Srama, R., Dikarev, V., Helfert, S., Moragas-Klostermeyer, G.,
  Krivov, A.~V., Srem{\v c}evi{\'c}, M., Tuzzolino, A.~J., Economou, T., and
  Gr{\"u}n, E. (2006).
\newblock {Cassini Dust Measurements at Enceladus and Implications for the
  Origin of the E Ring}.
\newblock {\em Science}, 311(5):1416--1418.

\bibitem[Sparks et~al., 2016]{Sparks:2016gl}
Sparks, W.~B., Hand, K.~P., McGrath, M.~A., Bergeron, E., Cracraft, M., and
  Deustua, S.~E. (2016).
\newblock {Probing for Evidence of Plumes on Europa with HST/STIS}.
\newblock {\em The Astrophysical Journal}, 829(2):1--21.

\bibitem[Spencer et~al., 2006]{2006Sci...311.1401S}
Spencer, J.~R., Pearl, J.~C., Segura, M., Flasar, F.~M., Mamoutkine, A.,
  Romani, P., Buratti, B.~J., Hendrix, A.~R., Spilker, L., and Lopes, R. M.~C.
  (2006).
\newblock {Cassini Encounters Enceladus: Background and the Discovery of a
  South Polar Hot Spot}.
\newblock {\em Science}, 311(5):1401--1405.

\bibitem[Spitale et~al., 2015]{2015Natur.521...57S}
Spitale, J.~N., Hurford, T.~A., Rhoden, A.~R., Berkson, E.~E., and Platts,
  S.~S. (2015).
\newblock {Curtain eruptions from Enceladus' south-polar terrain}.
\newblock {\em Nature}, 521(7):57--60.

\bibitem[Srama et~al., 2004]{Srama:2004uz}
Srama, R., Ahrens, T., Altobelli, N., Auer, S., Bradley, J., Burton, M.,
  Dikarev, V., Economou, T., Fechtig, H., Gorlich, M., Grande, M., Graps, A.,
  Gr{\"u}n, E., Havnes, O., Helfert, S., Hor{\'a}nyi, M., Igenbergs, E.,
  Jessberger, E., Johnson, T.~V., Kempf, S., Krivov, A.~V., Kr{\"u}ger, H.,
  Mocker-Ahlreep, A., Moragas-Klostermeyer, G., Lamy, P., Landgraf, M.,
  Linkert, D., Linkert, G., Lura, F., McDonnell, J., Mohlmann, D., Morfill,
  G.~E., Muller, M., Roy, M., Schafer, G., Schlotzhauer, G., Schwehm, G.,
  Spahn, F., Stubig, M., Svestka, J., Tschernjawski, V., Tuzzolino, A., Wasch,
  R., and Zook, H.~A. (2004).
\newblock {The Cassini Cosmic Dust Analyzer}.
\newblock {\em Space Science Reviews}, 114:465--518.

\bibitem[Srem{\v c}evi{\'c} et~al., 2003]{Sremcevic:2003gf}
Srem{\v c}evi{\'c}, M., Krivov, A.~V., and Spahn, F. (2003).
\newblock {Impact-generated dust clouds around planetary satellites: asymmetry
  effects}.
\newblock {\em Planetary and Space Science}, 51(7-8):455--471.

\bibitem[Strom et~al., 1979]{1979Natur.280..733S}
Strom, R.~G., Terrile, J.~R., Masursky, H., and Hansen, C. (1979).
\newblock {Volcanic eruption plumes on Io}.
\newblock {\em Nature}, 280:733 – 736.

\end{thebibliography}

\end{document}